\documentclass[10pt,a4paper,twocolumn,english,prb]{revtex4}
\usepackage[T1]{fontenc}
\usepackage[latin9]{inputenc}
\usepackage{color}
\usepackage{babel}
\usepackage{amstext}
\usepackage{graphicx}
\usepackage[unicode=true,pdfusetitle,
 bookmarks=true,bookmarksnumbered=false,bookmarksopen=false,
 breaklinks=false,pdfborder={0 0 1},backref=false,colorlinks=true]
 {hyperref}

\makeatletter

\pdfpageheight\paperheight
\pdfpagewidth\paperwidth

\@ifundefined{textcolor}{}
{%
 \definecolor{BLACK}{gray}{0}
 \definecolor{WHITE}{gray}{1}
 \definecolor{RED}{rgb}{1,0,0}
 \definecolor{GREEN}{rgb}{0,1,0}
 \definecolor{BLUE}{rgb}{0,0,1}
 \definecolor{CYAN}{cmyk}{1,0,0,0}
 \definecolor{MAGENTA}{cmyk}{0,1,0,0}
 \definecolor{YELLOW}{cmyk}{0,0,1,0}
}

\DeclareMathSizes{2}{2}{2}{2}
\renewcommand\[{\begin{equation}}
\renewcommand\]{\end{equation}}

\makeatother

\begin{document}

\title{Spectral function of the Higgs mode in $4-\varepsilon$ dimensions}

\author{Yaniv Tenenbaum Katan and Daniel Podolsky}

\affiliation{Physics Department, Technion -- Israel Institute of Technology, Haifa
32000, Israel}
\begin{abstract}
We investigate the amplitude (Higgs) mode of the relativistic $O\left(N\right)$
model in the vicinity of the Wilson-Fisher quantum critical point
in $D=4-\varepsilon$ spacetime dimensions. We compute the universal
part of the scalar spectral function near the transition, to leading
non-trivial order in the ordered phase, and to next to leading order
in both the disordered phase and the quantum critical regime. We find
that, in the disordered phase, the spectral function has a threshold
behavior with no Higgs-like peak, whereas in the ordered phase, the
Higgs mode appears as a well defined resonance. The pole associated
with this resonance is purely real in the $D\to3+1$ limit, evolving
smoothly with dimensionality to become purely imaginary at $D=2+1$
in the $N\to\infty$ limit. Our results complement previous studies
of the scalar spectral function, and demonstrate that the resonance
found in these studies can indeed be directly identified with the
Higgs mode.

\bigskip{}

\end{abstract}
\maketitle
\global\long\def\MD{\mathcal{D}}
 \global\long\def\MU{\mathcal{U}}
 {\tiny{}}\global\long\def\d{{\rm d}}
{\tiny{} }{\tiny \par}

\section{Introduction}

Relativistic field theories with $O\left(N\right)$ symmetry describe
spontaneous symmetry breaking phase transitions in a large variety
of quantum systems. In condensed matter and cold atomic systems, the
experimental realizations of these models include, for example, the
transverse field Ising model ($N=1$) \cite{coldea2010quantum,sachdev2002mottinsulators,simon2011quantum,ZeemanLadders},
the superfluid to Mott insulator transition ($N=2$) \cite{spielman2007mottinsulator,endres2012thehiggs},
and the Néel transition in dimerized antiferromagnets ($N=3$) \cite{chow1998singular,ruegg2008quantum}.

Breaking of symmetry gives rise to collective modes. When the broken
symmetry is continuous, these excitations include massless Goldstone
modes, which are related to fluctuations in the direction of the order
parameter, and, in systems with emergent relativistic invariance,
to a massive Higgs mode associated with fluctuations in the order
parameter amplitude\cite{varma2002,pekker2014amplitude}.

The Higgs mode and the Goldstone modes are both long lived at the
mean field level. However, effects beyond mean field allow for the
decay of the Higgs mode into pairs of Goldstone modes. As a result,
the Higgs mode acquires a finite lifetime, thus bringing into question
its visibility in experiments. This question is especially relevant
to systems in $D<4$ spacetime dimensions, that is, in $d<3$ spatial
dimensions (we consider relativistically invariant quantum critical
points, for which $D=d+1$). In this case the emission of Goldstone
modes leads to an infrared divergence in the longitudinal susceptibility
\cite{sachdev1999universal,zwerger2004anomalous,dupuis2011infrared},
the standard correlation function used to probe the Higgs mode. However,
this effect is sensitive to the response function used to probe the
mode, and in particular to its symmetry \cite{chubukov1994theoryof,lindner2010conductivity,podolsky2011visibility}
. Specifically, such infrared divergence has been shown to be absent
the scalar response function \cite{podolsky2011visibility}. A measurement
of this type has been performed on cold bosons in an optical lattice,
where the Higgs mode was experimentally observed near the Mott insulator
to superfluid transition \cite{endres2012thehiggs}.

The spacetime dimensionality $D$ also plays an important role in
the nature of the ordering transition at the quantum critical point
(QCP). For $D=3+1$, the non-linear coupling decays logarithmically
as the QCP is approached and the QCP itself is a Gaussian fixed point\cite{wilson1972critical,kardar2007statistical}.
This leads to a Higgs decay width that tends to zero faster than its
mass, rendering the Higgs mode ``critically-well defined'' \cite{affleck1992longitudinal,sachdev1999universal,oitmaa2012universal}.
By contrast, for $2<D<4$ the QCP is a Wilson-Fisher fixed point and
the interactions remain finite as the Higgs mass approaches zero.
In this sense, the interactions are fundamentally strong close to
the critical point, and it is therefore interesting to study the nature
of the Higgs resonance in this case. 

The Higgs mode near the QCP at $D=3$ has been studied using a variety
of methods. In particular, the scalar susceptibility was computed
analytically in the large $N$ limit \cite{podolsky2012spectral},
numerically in Quantum Monte Carlo (QMC) simulations \cite{pollet2012higgsmode,chen2013universal,gazit2013fateof,gazit2013dynamics},
and through a non-perturbative renormalization group (NPRG) approach
\cite{RanconDupuis}. These methods provide valuable information on
the nature of the Higgs mode near the QCP, but each approach has its
limitations. The analytic results may be far from experimentally relevant
systems, for which $N\le3$. Similarly, the QMC and NPRG methods rely
on numerical analytic continuation of Matsubara frequency response
functions, a procedure that is difficult to control. As a result,
some disagreement still exists regarding various properties of the
Higgs mode. These include quantitative questions such as the precise
value of the mass of the Higgs mode, as well as qualitative questions
such as the possible appearance of a Higgs-like resonance in the disordered
phase.

In this paper we address these questions following a different approach.
We obtain the scalar susceptibility near the Wilson-Fisher fixed point
near $D=3+1$ spacetime dimensions, using a $D=4-\varepsilon$ calculation
that is controlled for small $\varepsilon$. We then study the nature
of the Higgs excitation by extracting the universal component of the
scalar response function near the QCP. This approach has its own limitations
when applied to $D=2+1$; however, it provides valuable information
that complements previous approaches.

The main results of our analysis are: (1)\ In the disordered phase,
we find that the scalar spectral function has a threshold behavior
without an accompanying Higgs-like resonance. This contrasts numerical
results in $D=2+1$ \cite{chen2013universal}. (2)\ In the ordered
phase, we find that the scalar spectral function features a sharp
Higgs peak and extract its associated pole near the Wilson-Fisher
fixed point. Furthermore, in the $N\rightarrow\infty$ limit, we study
the position of the pole as a function of space-time dimensionality,
and find that it evolves smoothly from a purely real pole at $D=3+1$
to a purely imaginary pole at $D=2+1$. The analytic structure of
the scalar spectral function was previously studied in the large $N$
approximation \cite{podolsky2012spectral} and by holographic methods
\cite{HolographicSuperfluids}. In particular, in Ref. \cite{podolsky2012spectral}
it was shown that for large but finite $N$, the pole at $D=2+1$
picks up a small real component. Our results show that this pole is
indeed smoothly connected to the sharp Higgs peak at $D=3+1$. (3)\ In
the quantum critical regime, we find that the universal scaling function
has a peak at finite frequencies near $D=3+1$. This peak disappears
as one approaches $D=2+1$.

This article is organized in follows. In Sec.\ \ref{sec:General-Formalism}
we briefly review the $O\left(N\right)$ model, the physical observables
studied in this work, and their expected scaling near the QCP. In
Sec.\ \ref{sec:The-Symmetric-Phase} we focus on the disordered phase;
we calculate the single particle gap, the scalar susceptibility, extract
the universal scaling function to next to leading order in $\varepsilon$,
and discuss its properties. In Sec.\ \ref{sec:The-Ordered-Phase}
we extend the analysis to the ordered phase, and evaluate the scalar
susceptibility and its universal scaling function to the first non-trivial
order in $\varepsilon$. In addition, we evaluate the scalar response
function for $N=\infty$ and general $D$. In Sec.\ \ref{sec:Expansion-in-Finite}
we compute the universal scaling function in the quantum critical
regime, to next to leading order. In Sec.\ \ref{sec:Summary-and-Conclusions}
we provide a summary and conclusions. In Appendix \ref{sec:Suscpetibilities Goldstone}
we present the detailed calculations in the ordered phase. In Appendix
\ref{appendix:COMPUTATION-OF-} we compute the polarization bubble
in the quantum critical regime.

\section{General Formalism\label{sec:General-Formalism}}

\subsection{Model}

We study the partition function defined by the path integral

\begin{equation}
\mathcal{Z}\begin{array}{c}
=\int\MD\phi_{\alpha}\exp\left(-S\left[\phi\right]\right)\end{array}\label{eq:Zu=00005BJ=00005D,phi}
\end{equation}
\begin{equation}
S\left[\phi\right]=\int_{x}\left[\frac{1}{2}\left(\partial_{\mu}\phi_{\alpha}\left(x\right)\right)^{2}+\frac{r}{2}\phi_{\alpha}^{2}\left(x\right)+\frac{U}{8}\left(\phi_{\alpha}^{2}\left(x\right)\right)^{2}\right].\label{eq:Action}
\end{equation}
Here, $\phi_{\alpha}$ is a real field with $N$ components, $\alpha\in\left\{ 1...N\right\} $.
The action is defined on Euclidean space-time with $\int_{x}=\int d^{D}x$
where $D$ is the space-time dimension. The action in Eq.\ (\ref{eq:Action})
has relativistic invariance, in which the coordinates have been scaled
such that the speed of sound is one.

The system described by Eq.\ (\ref{eq:Zu=00005BJ=00005D,phi}) undergoes
a quantum phase transition at a critical value $r_{c}$ \cite{fisher1974scaling,nelson1976twopoint,kardar2007statistical}.
For $r>r_{c}$, the system is in a disordered phase with $\left\langle \phi\right\rangle =0$.
In this phase the $O\left(N\right)$ symmetry is preserved and there
are $N$ degenerate gapped modes. By contrast, for $r<r_{c}$ the
system is in an ordered phase and the $\phi$ field acquires an expectation
value (EV), $\left\langle \phi\right\rangle =\left(\phi_{0},0,....\right)$
which breaks the $O\left(N\right)$ symmetry down to $O\left(N-1\right)$. 

The breaking of the $O\left(N\right)$ symmetry leads to the emergence
of collective modes which correspond to fluctuations of the order
parameter, 
\[
\phi=\left(\phi_{0}+\sigma,\vec{\pi}\right)
\]
where the $N-1$ component field $\vec{\pi}$ corresponds to the $N-1$
gapless Goldstone modes and the scalar field $\sigma$ is associated
with the Higgs mode.

\subsection{Physical Observables}

The two-point tensor dynamical correlation function of the $\phi_{\alpha}$
field is defined by,

\begin{equation}
\begin{array}{c}
\chi_{\alpha\beta}\left(p\right)=\int_{x}e^{-i\mathbf{p}\cdot\mathbf{x}}\left[\left\langle \phi_{\alpha}\left(x\right)\phi_{\beta}\left(0\right)\right\rangle -\left\langle \phi_{\alpha}\left(x\right)\right\rangle \left\langle \phi_{\beta}\left(0\right)\right\rangle \right].\end{array}\label{eq:Chi lONG}
\end{equation}
In the ordered phase, the amplitude fluctuations of the order parameter
can be probed by the longitudinal susceptibility,
\begin{equation}
\chi_{11}\left(p\right)=\int_{x}e^{-i\mathbf{p}\cdot\mathbf{x}}\left[\left\langle \sigma\left(x\right)\sigma\left(0\right)\right\rangle -\left\langle \sigma\left(x\right)\right\rangle \left\langle \sigma\left(0\right)\right\rangle \right]\label{eq:LOngitudinal}
\end{equation}
which is the two-point correlation function of the $\sigma$ field.

Within the mean field approximation, Eq.\ (\ref{eq:LOngitudinal})
has a pole corresponding to a gapped excitation \cite{sachdev1999universal},
identified as the Higgs mode. However beyond mean field level, the
peak of the Higgs mode broadens as a result of the decay of the Higgs
into pairs of Goldstone modes. In particular, for $D<4$, the longitudinal
susceptibility has divergent spectral weight at low frequencies, which
overwhelms the Higgs resonance close to the critical point\cite{sachdev1999universal,zwerger2004anomalous}.

We will focus instead on a second observable, the scalar susceptibility,
which is the two-point correlation function of the amplitude squared
of the field $\phi$,

\begin{equation}
\begin{array}{c}
\chi_{_{s}}\left(p\right)=\end{array}\int_{x}e^{-i\mathbf{p}\cdot\mathbf{x}}\left[\left\langle \phi_{\alpha}^{2}\left(x\right)\phi_{\beta}^{2}\left(0\right)\right\rangle -\left\langle \phi_{\alpha}^{2}\left(x\right)\right\rangle \left\langle \phi_{\beta}^{2}\left(0\right)\right\rangle \right].\label{eq:S(P) definition}
\end{equation}
It has been argued \cite{podolsky2011visibility} that the scalar
susceptibility is less sensitive to the emission of Goldstone modes,
and therefore produces a sharper resonance for the Higgs mass. 

We are interested in the dynamical scalar structure factor function
$S\left(\omega\right)$, obtained by analytic continuation of the
scalar susceptibility, 
\begin{equation}
S\left(\omega\right)=\Im\left\{ \chi_{s}\left(p\rightarrow-i\omega+0^{+}\right)\right\} \label{eq:S(w) definition}
\end{equation}
Equation\ (\ref{eq:S(w) definition}) corresponds to zero momentum
and a finite probe frequency $\omega$.

\subsection{The Wilson-Fisher Fixed Point}

The modern description of critical behavior is based on the assumption
that near phase transitions, the long distance properties of a system
are determined by the large correlation length $\xi$, which is the
only important length scale \cite{kadanoff1966scaling,migdalrecursion}.
In particular, the critical behavior is dominated by fluctuations
that are self similar up to the scale of the correlation length $\xi$.
This last property can be used to build a description of the critical
behavior through the Renormalization Group (RG). The RG procedure
consists of gradually eliminating the correlated degrees of freedom
at length scales $x\ll\xi$, until only uncorrelated, simple degrees
of freedom remain at the length scale $\xi$. In this process, different
fixed points of the RG procedure correspond to either phases or to
phase transitions.

Applying the RG procedure to Eq.\ (\ref{eq:Zu=00005BJ=00005D,phi})
yields that at $D=4-\varepsilon$ space time dimensions, this system
undergoes a phase transition described by the Wilson-Fisher fixed
point \cite{wilson1972critical,fisher1974scaling,nelson1976twopoint,kardar2007statistical}.
This result is derived for $D=4-\varepsilon$ space time dimensions,
but it is generally believed that the fixed point can be smoothly
evolved to describe the phase transition in the range $2<D<4$. 

For $D=4-\varepsilon$, the interaction coupling $U$ at the Wilson-Fisher
fixed point is of $\mathcal{O}\left(\varepsilon\right)$. To $\mathcal{O}\left(\varepsilon^{2}\right)$,
\cite{fisher1974scaling}
\begin{equation}
U_{c}=\frac{2\varepsilon}{N+8}\frac{1}{\Omega_{\varepsilon}}\left(1+\frac{9N+42}{\left(N+8\right)^{2}}\varepsilon\right)\label{eq:Uc}
\end{equation}
where $\Omega_{\varepsilon}=\Lambda^{-\varepsilon}K_{4-\varepsilon}$
, $K_{D}=\frac{2^{1-D}\pi^{-D/2}}{\Gamma\left(\frac{D}{2}\right)}$
is the normalized area of a $D$-dimensional sphere, $\Lambda$ is
the implicit ultraviolet cutoff, and $\Gamma$ is Euler's Gamma function.
This allows for a controlled expansion of the physical observables
in powers of $\varepsilon$.

In the case of the Wilson-Fisher point, deviations of $U$ away from
$U_{c}$ are irrelevant and flow only slowly to zero. Therefore, in
order to extract universal properties near the critical point, we
will set $U=U_{c}$ and use $\delta r=r-r_{c}$ as the tuning parameter
across the transition \cite{fisher1974scaling,nelson1976twopoint,kardar2007statistical}.

\subsection{Scaling\label{sub:Scaling}}

We proceed to obtain the scaling limit \cite{podolsky2012spectral}
of the dynamical scalar structure factor $S\left(\omega\right)$ near
the phase transition, in terms of the critical exponents. These critical
exponents can be obtained from the RG procedure at the Wilson-Fisher
fixed point. We start by deriving the free energy density of Eq.\ (\ref{eq:Zu=00005BJ=00005D,phi}),
$f=-\frac{1}{V}\ln Z$, with respect to $r$, 
\[
\begin{array}{c}
\frac{\partial^{2}f}{\partial r^{2}}=\frac{1}{4V}\int_{x,y}\left[\left\langle \phi_{\alpha}^{2}\left(x\right)\phi_{\beta}^{2}\left(y\right)\right\rangle -\left\langle \phi_{\alpha}^{2}\left(x\right)\right\rangle \left\langle \phi_{\beta}^{2}\left(y\right)\right\rangle \right]-\frac{1}{U}\end{array}
\]
which is the scalar susceptibility for $p=0$, up to additive and
multiplicative constants. $f$ can be written as a sum of regular
and singular parts, where according to the hyperscaling hypothesis
the latter scales as $\xi^{-D}$. The correlation length, $\xi$ satisfies
$\xi\propto\left|r-r_{c}\right|^{-\nu}$, where $\nu$ is the correlation
length exponent. It follows that 
\[
S\left(\omega=0\right)\propto\xi^{-D+2/\nu}+regular\: part.
\]
For a relativistic theory, $\xi\propto\frac{1}{\Delta}$ where $\Delta$
is the energy gap of single particle excitations in the disordered
phase. $\Delta$ serves as a characteristic energy scale at the ordered
phase. We conclude that 
\begin{equation}
S\left(\omega\right)=\mathcal{A}_{\pm}\Delta^{2\alpha}\Phi_{\pm}\left(\frac{\omega}{\Delta}\right)+regular\: part\label{eq:S=00005Bw=00005D Scale}
\end{equation}
where $2\alpha=D-\frac{2}{\nu}$. The regular part is non-universal
and is analytic in $\delta r$, and$\Phi_{\pm}$ are universal scaling
functions which describe the critical behavior of $S\left(\omega\right)$
in the disordered $\left(\Phi_{+}\right)$ and ordered $\left(\Phi_{-}\right)$
phases.

In order to fix the overall amplitudes $\mathcal{A}_{\pm}$, we look
at the asymptotic behavior of $S(\omega)$ in different regimes. In
the disordered phase, we find that $S(\omega)$ has a threshold at
$\omega=2\Delta$. Near the threshold,

\begin{equation}
S(\omega)\sim\mathcal{A}_{+}\Delta^{2\alpha}\left(\frac{\delta\omega}{\Delta}\right)^{(D-3)/2}\Theta(\delta\omega)\,,\label{eq:AplusDef}
\end{equation}
where $\delta\omega\equiv\omega-2\Delta$ and where $\Theta\left(x\right)$
is the Heaviside step function. On the other hand, in the ordered
phase and at low frequencies $0<\omega\ll\Delta$, we find

\begin{equation}
S(\omega)\sim\mathcal{A}_{-}\Delta^{2\alpha}\left(\frac{\omega}{\Delta}\right)^{D}\:.\label{eq:AminusDef}
\end{equation}
We will use these asymptotic forms to define $\mathcal{A}_{\pm}$.
Their individual values are not universal (\emph{e.g.} they depend
on the UV cutoff), but their ratio is. We find,

\begin{equation}
\frac{\mathcal{A}_{+}}{\mathcal{A}_{-}}=\frac{4N}{N-1}\label{eq:Arat}
\end{equation}
Note that the definition for $\mathcal{A}_{+}$ is slightly different
from that used in Ref. \cite{podolsky2012spectral}, where the logarithmic
threshold singularity, specific to $D=2+1$, was used. Hence, a straightforward
comparison of the ratio obtained in both approaches is not possible. 

At $T>0$, the temperature serves as a second characteristic energy
scale. In particular, at $r=r_{c}$, the scaling of $S\left(\omega\right)$
at finite temperatures becomes 
\begin{equation}
S\left(\omega\right)=\mathcal{A}_{T}T^{2\alpha}\Phi_{T}\left(\frac{\omega}{T}\right)+regular\: part\label{eq:Scaling T>0}
\end{equation}
where $\Phi_{T}$ is the thermal universal scaling function.

We will compute the universal scaling functions $\Phi_{\pm}$ and
$\Phi_{T}$ near the Wilson-Fisher fixed point.

\section{The Disordered Phase \label{sec:The-Symmetric-Phase}}

In this section we focus on the disordered phase $\left(\delta r>0\right)$
at $T=0$, in which the $O\left(N\right)$ symmetry is preserved.
We extract the single-particle gap $\Delta$ from the poles of the
longitudinal susceptibility, obtain the universal part of the scalar
susceptibility $\Phi_{+}\left(\frac{\omega}{\Delta}\right)$, and
study the threshold singularity of this physical observable.

\subsection{Critical Point}

We begin by determining the value of $r$ at the transition, by requiring
\cite{podolsky2012spectral} $G\left(0,r_{c}\right)^{-1}=0$, where
$G\left(p,r\right)=\chi_{\alpha\alpha}\left(p,r\right)$ is the two
point correlation function of the $\phi_{\alpha}$ field in the disordered
phase. 

We obtain $r_{c}$ for $D=4-\varepsilon$ at $U=U_{c}$, as given
in Eq.\ (\ref{eq:Uc}). The Dyson expansion for $G\left(p,r\right)$
to $\mathcal{O}\left(\varepsilon^{2}\right)$ yields
\begin{eqnarray}
G^{-1}\left(p,r\right) & = & p^{2}+r+\frac{U_{c}\left(N+2\right)}{2}\int_{q}\frac{1}{q^{2}+r-\Sigma_{1}\left(r\right)}\nonumber \\
 &  & -\frac{U_{c}^{2}\left(N+2\right)}{2}\int_{q}\frac{\Pi\left(q,\sqrt{r}\right)}{\left(\mathbf{p}+\mathbf{q}\right)^{2}+r}\label{eq:phi propagator}
\end{eqnarray}
where $\int_{q}\equiv\int\frac{d^{4-\varepsilon}q}{\left(2\pi\right)^{4-\varepsilon}}$.
Here, $\Pi\left(q,\sqrt{r}\right)$ is the polarization bubble,
\begin{equation}
\Pi\left(q,\sqrt{r}\right)=\int_{k}\frac{1}{\left(\left(\mathbf{q}+\mathbf{k}\right)^{2}+r\right)\left(k^{2}+r\right)},\label{eq:pi(p,r) symbolic}
\end{equation}
and
\begin{equation}
\Sigma_{1}\left(r\right)=-\frac{U_{c}\left(N+2\right)}{2}\int_{q}\frac{1}{q^{2}+r}.\label{eq:selfenergy}
\end{equation}
The equation for $r_{c}$ is then 
\begin{eqnarray}
0 & = & r_{c}+\frac{U_{c}\left(N+2\right)}{2}\int_{q}\frac{1}{q^{2}+r_{c}-\Sigma_{1}\left(r_{c}\right)}\label{eq:Equation for rc}\\
 &  & -\frac{U_{c}^{2}\left(N+2\right)}{2}\int_{q}\frac{\Pi\left(q,\sqrt{r_{c}}\right)}{q^{2}+r_{c}}.\nonumber 
\end{eqnarray}
From inspection of Eqs.\ (\ref{eq:selfenergy}) and (\ref{eq:Equation for rc}),
we conclude that $r_{c}=\Sigma_{1}\left(r_{c}\right)+\mathcal{O}\left(\varepsilon^{2}\right)$.
We can use this to write Eq.\ (\ref{eq:Equation for rc}) to $\mathcal{O}\left(\varepsilon^{2}\right)$
as 
\begin{eqnarray}
r_{c} & = & -\frac{U_{c}\left(N+2\right)}{2}\left(\int_{q}\frac{1}{q^{2}}-U_{c}\int_{q}\frac{\Pi\left(q,0\right)}{q^{2}}\right).\label{eq:Gff(p=00003D0)}
\end{eqnarray}
By performing the integrals in Eq.\ (\ref{eq:Gff(p=00003D0)}), we
find
\begin{eqnarray}
r_{c} & = & \frac{U_{c}\left(N+2\right)K_{D}}{2}\frac{\Lambda^{D-2}}{2-D}\left(1-\frac{U_{c}K_{D}\Lambda^{D-4}}{D-3}\right).\label{eq:rc, general dimension}
\end{eqnarray}
Hence, using Eq.\ (\ref{eq:Uc}), we obtain
\begin{equation}
r_{c}=-\varepsilon\frac{\Lambda^{2-2\varepsilon}}{2}\frac{N+2}{N+8}\left(1-\varepsilon\frac{N^{2}+30N+116}{2(N+8)^{2}}\right)+\mathcal{O}\left(\varepsilon^{3}\right).\label{eq:rc}
\end{equation}
In what follows we write 
\begin{equation}
r=r_{c}+\delta r\label{eq:deltar}
\end{equation}
such that the quantum phase transition occurs at $\delta r=0$.

\subsection{Single Particle Gap\label{sub:Single-Particle-Gap}}

The single particle gap $\Delta$ is determined by the condition\cite{podolsky2012spectral}
$G\left(p=-i\Delta,\delta r\right)^{-1}=0$. Using Eqs.\ (\ref{eq:selfenergy})
and (\ref{eq:rc}), Eq.\ (\ref{eq:phi propagator}) becomes
\begin{eqnarray*}
G^{-1}\left(p,\delta r\right) & = & p^{2}+\delta r+\frac{U_{c}\left(N+2\right)}{2}\times\\
 &  & \left(\begin{array}{c}
\int_{q}\frac{1}{q^{2}+\delta r}\end{array}-\begin{array}{c}
\int_{q}\frac{1}{q^{2}}\end{array}\right)+\mathcal{O}\left(\varepsilon^{2}\right).
\end{eqnarray*}
 It follows that the pole of $G\left(p,\delta r\right)$ is at $p^{2}=-\Delta^{2}$,
where 
\begin{eqnarray}
\Delta^{2} & = & \delta r\left(1-\frac{\varepsilon}{2}\frac{N+2}{N+8}\ln\frac{\Lambda^{2}}{\delta r}\right)+\mathcal{O}\left(\varepsilon^{2}\right)\label{eq:Symmetric phase pole}\\
 & = & \Lambda^{2}\left(\frac{\delta r}{\Lambda^{2}}\right){}^{2\nu}+\mathcal{O}\left(\varepsilon^{2}\right),\nonumber 
\end{eqnarray}
and $\nu=\frac{1}{2}+\frac{\varepsilon}{4}\frac{N+2}{N+8}$. Equation
(\ref{eq:Symmetric phase pole}) provides the required relation between
the energy gap $\Delta$ and the tuning parameter $\delta r$\cite{wilson1972critical,fisher1973scaling,sachdev1999universal,kardar2007statistical,podolsky2012spectral}.
Note that to this order in $\varepsilon$, the renormalized propagator
is simply

\begin{equation}
G\left(p,\delta r\right)=\frac{1}{p^{2}+\Delta^{2}}+\mathcal{O}(\varepsilon^{2})\label{eq:renormG}
\end{equation}
 where $\Delta$ is related to $\delta r$ by Eq.\ (\ref{eq:Symmetric phase pole}).

\subsection{Scalar Susceptibility\label{sub:Scalar-Susceptibility}}

\begin{figure}
\includegraphics[width=0.75\columnwidth]{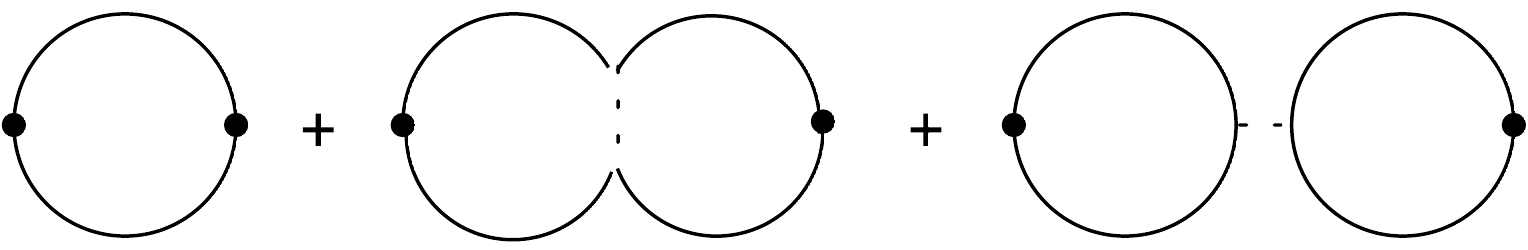} 

\protect\caption{The diagrams which contribute to the scalar susceptibility in the
disordered phase, to first non-trivial order. The solid lines are
renormalized propagators, Eq.\ (\ref{eq:renormG}). \label{fig:S(p) symmetric diagrams}}
\end{figure}

The scalar susceptibility can be evaluated order by order in $\varepsilon$.
The diagrammatic expansion of this calculation is given to $\mathcal{O}\left(\varepsilon\right)$
in Fig.\ \ref{fig:S(p) symmetric diagrams}. We find 
\begin{equation}
\chi_{s}\left(p,\delta r\right)=2N\Pi\left(p,\Delta\right)-U_{c}N\left(N+2\right)\Pi\left(p,\Delta\right)^{2}+\mathcal{O}\left(\varepsilon^{2}\right).\label{eq:S(p) symmetric scehmatic}
\end{equation}
Here,$\Pi\left(p,\Delta\right)$ is the polarization bubble, Eq.\ (\ref{eq:pi(p,r) symbolic}),
which evaluates to
\begin{eqnarray}
\Pi\left(p,\Delta\right) & = & \Omega_{\varepsilon}\left[\frac{1}{2}-\frac{\tanh^{-1}x}{x}+\frac{1}{2}\ln\frac{\Lambda^{2}}{\Delta^{2}}\right.\label{eq:pi(p,r)}\\
 &  & +\varepsilon\left\{ \frac{1}{4x}\left(\text{Li}_{2}\left(\frac{1+x}{2}\right)-\text{Li}_{2}\left(\frac{1+x}{2}\right)\right)\right.\nonumber \\
 &  & -\frac{\tanh^{-1}x}{2x}\left(1-\frac{1}{2}\ln\frac{p^{2}}{x^{2}\Delta^{2}}-\ln\frac{\Lambda^{2}}{\Delta^{2}}\right)\nonumber \\
 &  & \left.\left.+\frac{3}{8}+\frac{\pi^{2}}{6}+\frac{1}{8}\left(1+\ln\frac{\Lambda^{2}}{\Delta^{2}}\right)^{2}\right\} \right]+\mathcal{O}\left(\varepsilon^{2}\right),\nonumber 
\end{eqnarray}
where $x=\frac{p}{\sqrt{p^{2}+4\Delta^{2}}}$ and $Li_{2}\left(z\right)$
is the dilogarithm function, defined by $\text{Li}_{2}\left(z\right)=-\int_{0}^{z}\frac{\ln\left(1-t\right)}{t}dt$. 

The explicit expression for the scalar susceptibility is obtained
by inserting Eq.\ (\ref{eq:pi(p,r)}) and Eq.\ (\ref{eq:Uc}) into
Eq.\ (\ref{eq:S(p) symmetric scehmatic}), 
\begin{eqnarray}
\frac{\Omega_{\varepsilon}^{-1}}{N}\chi_{s}\left(p,\delta r\right) & = & 1+\ln\frac{\Lambda^{2}}{\Delta^{2}}-\frac{\alpha_{1}\varepsilon}{2}\left(1+\ln\frac{\Lambda^{2}}{\Delta^{2}}\right)^{2}+\frac{\alpha_{1}\varepsilon}{2}\nonumber \\
 &  & -\frac{2\tanh^{-1}x}{x}\left[1-\alpha_{1}\varepsilon\left(1+\ln\frac{\Lambda^{2}}{\Delta^{2}}\right)\right]\nonumber \\
 &  & +\varepsilon\chi_{2}(p)+\mathcal{O}\left(\varepsilon^{2}\right)\label{eq:S(p) Symmetric}
\end{eqnarray}
where
\begin{eqnarray*}
\chi_{2}\left(p\right) & = & \frac{\pi^{2}}{12}+\frac{\tanh^{-1}x}{2x}\ln\frac{p^{2}}{x^{2}\Delta^{2}}-2\frac{N+2}{N+8}\left(\frac{\tanh^{-1}x}{x}\right)^{2}\\
 &  & +\frac{1}{2x}\left[\text{Li}_{2}\left(\frac{1-x}{2}\right)-\text{Li}_{2}\left(\frac{1+x}{2}\right)\right]+\frac{1}{2}\frac{N+14}{N+8},
\end{eqnarray*}
and $\alpha_{1}$ is given in Eq.\ (\ref{eq:alpha1,2}) below.

\subsection{Universal Scaling Function \label{Appendix:Universal part Symmetric}}

As argued in Sec.\ \ref{sub:Scaling}, near the transition, the scalar
susceptibility takes the form,
\begin{equation}
\chi_{s}\left(p,\delta r\right)=\mathcal{A}_{+}\Delta^{2\alpha}\tilde{\Phi}_{+}\left(\frac{p}{\Delta}\right)+regular\: part\label{eq:S=00005Bp=00005D scaling H}
\end{equation}
where $2\alpha=D-2/\nu.$ To $\mathcal{O}(\varepsilon^{2}),$ the
critical exponent $\alpha$ is given by\cite{fisher1974scaling}
\[
\alpha=\alpha_{1}\varepsilon+\alpha_{2}\varepsilon^{2}+\mathcal{O}\left(\varepsilon^{3}\right),
\]
where,

\begin{eqnarray}
\alpha_{1} & = & \frac{1}{2}\frac{N-4}{N+8},\label{eq:alpha1,2}\\
\alpha_{2} & = & \frac{\left(N+2\right)\left(13N+44\right)}{2(N+8)^{3}}.\nonumber 
\end{eqnarray}
$\tilde{\Phi}_{+}\left(\frac{\omega}{\Delta}\right)$ is related to
the universal scaling function $\Phi_{+}\left(\frac{\omega}{\Delta}\right)$
by a Wick rotation, 
\begin{equation}
\Phi_{+}\left(\frac{\omega}{\Delta}\right)=\Im\left\{ \tilde{\Phi}_{+}\left(-\frac{i\omega}{\Delta}+0^{+}\right)\right\} .\label{eq:Phi and H}
\end{equation}

We evaluate $\tilde{\Phi}_{+}\left(\frac{\omega}{\Delta}\right)$,
the universal component of Eq.\ (\ref{eq:S(p) Symmetric}). We choose
\begin{equation}
\mathcal{A}_{+}=\Omega_{\varepsilon}\Lambda^{-2\alpha}N\pi(1-\varepsilon\alpha_{1}-\frac{\varepsilon}{2}\ln2).\label{eq:Aplus}
\end{equation}
which gives the normalization in Eq.\ (\ref{eq:AplusDef}), as will
be shown later. Then, we can expand $\tilde{\Phi}_{+}\left(\frac{p}{\Delta}\right)$
in non-negative powers of $\varepsilon$ as $\tilde{\Phi}_{+}\left(\frac{p}{\Delta}\right)=\tilde{\Phi}_{0}+\varepsilon\tilde{\Phi}_{1}+\varepsilon^{2}\tilde{\Phi}_{2}+\mathcal{O}\left(\varepsilon^{3}\right)$
and write
\begin{eqnarray}
\mathcal{A}_{+}\Delta^{2\alpha}\tilde{\Phi}_{+} & = & \frac{\tilde{\Phi}_{0}}{\varepsilon}+\tilde{\Phi}_{1}-\alpha_{1}\tilde{\Phi}_{0}\lambda+\varepsilon\tilde{\Phi}_{2}\label{eq:dis_expansion}\\
 &  & -\left(\alpha_{2}\tilde{\Phi}_{0}+\alpha_{1}\tilde{\Phi}_{1}\right)\varepsilon\lambda\nonumber \\
 &  & +\frac{1}{2}\alpha_{1}^{2}\tilde{\Phi}_{0}\varepsilon\lambda^{2}+\mathcal{O}\left(\varepsilon^{2}\right)\nonumber 
\end{eqnarray}
where $\lambda\equiv\ln\frac{\Lambda^{2}}{\Delta^{2}}$. We can now
obtain $\tilde{\Phi}_{0},\tilde{\Phi}_{1}$ and $\tilde{\Phi}_{2}$
by comparing Eqs.\ (\ref{eq:S(p) Symmetric}) and (\ref{eq:dis_expansion})
order by order in both $\varepsilon$ and $\lambda$. Indeed, we find
that the scalar susceptibility is of the scaling form, Eq.\ (\ref{eq:S=00005Bp=00005D scaling H}),
with 
\begin{eqnarray}
\tilde{\Phi}_{+}\left(\frac{p}{\Delta}\right) & = & -\frac{2}{\pi}\left(1-\alpha_{1}\varepsilon\right)\frac{\tanh^{-1}x}{x}\label{eq:phit_minus}\\
 &  & +\frac{\varepsilon}{\pi}\chi_{2}\left(p\right)-\mathcal{C}+\mathcal{O}\left(\varepsilon^{2}\right)\nonumber 
\end{eqnarray}
where $\mathcal{C}=\frac{1}{\alpha_{1}\pi}\left(\frac{1}{\varepsilon}-\alpha_{1}-\frac{\alpha_{2}}{\alpha_{1}}\right)$
is a real constant which does not contribute to the real frequency
function $\Phi_{+}$. In addition, we find that the non-universal
part in Eq.\ (\ref{eq:S=00005Bp=00005D scaling H}) is given by the
constant 

\begin{equation}
\chi_{reg}=N\Omega_{\varepsilon}\left(\frac{1}{\alpha_{1}\varepsilon}-\frac{\alpha_{2}}{\alpha_{1}^{2}}\right).\label{eq:chireg}
\end{equation}

As a consistency check, we examine the asymptotic behavior of Eq.\ (\ref{eq:phit_minus})
in the limit $p\gg\Delta$. In this limit, the universal component
is expected to take the form \cite{podolsky2012spectral} 
\begin{equation}
\tilde{\Phi}_{+}\left(\frac{p}{\Delta}\right)\propto\left(\frac{p}{\Delta}\right)^{2\alpha}\label{eq:hyper_scale_limit}
\end{equation}
In the same limit, Eq.\ (\ref{eq:phit_minus}) becomes
\begin{eqnarray}
\tilde{\Phi}_{+}\left(\frac{p}{\Delta}\right) & \to & -\mathcal{C}-\frac{2}{\pi}\left(1-\alpha_{1}\varepsilon\right)\ln\frac{p}{\Delta}+\frac{\varepsilon}{2\pi}\frac{N+14}{N+8}\nonumber \\
 &  & -\frac{2}{\pi}\alpha_{1}\varepsilon\ln^{2}\frac{p}{\Delta}+\mathcal{O}\left(\frac{\Delta}{p}\right)+\mathcal{O}\left(\varepsilon^{2}\right)\nonumber \\
 & = & -\mathcal{C}\left(\frac{p}{\Delta}\right){}^{2\alpha}+\varepsilon\frac{1}{2\pi}\frac{N+14}{N+8}\nonumber \\
 &  & +\mathcal{O}\left(\frac{\Delta}{p}\right)+\mathcal{O}\left(\varepsilon^{2}\right),\label{eq:Chis_scalar_limit}
\end{eqnarray}
as expected. Here, we've used the relation $x^{y}=1+y\ln x+(y\ln x)^{2}/2+\mathcal{O}(y^{3})$.
Note that for $N=4$, the final expression is not well defined since
$\mathcal{C}=\infty$. This is an artifact of our working order in
$\varepsilon$ since, for $N=4$, $\alpha$ vanishes to$\mathcal{O}(\varepsilon)$,
and hence it is not possible to exponentiate the expression in the
top line. 

\begin{figure}
\includegraphics[width=1\columnwidth]{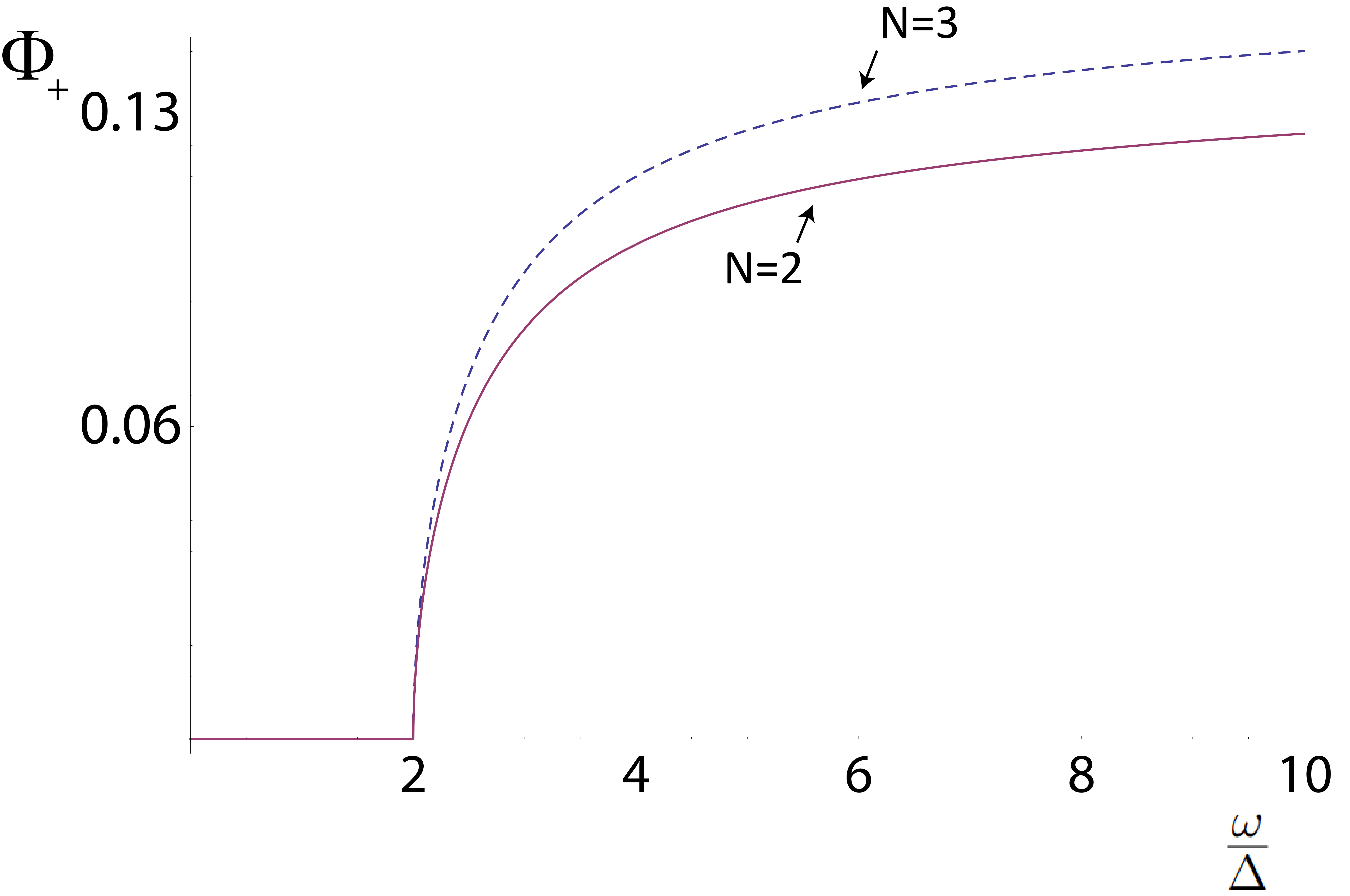}

\protect\caption{The universal scaling function in the disordered phase, $\Phi_{+}\left(\frac{\omega}{\Delta}\right)$,
to second order in $\varepsilon$. Results are for $\varepsilon=0.1$
and for $N=2,3$. We find that to this order, the shape of the universal
scaling function has very little dependence on $\varepsilon$. \label{fig:Im=00007Bphi=00007D symmetric}}
\end{figure}

The universal scaling function in the disordered phase can now be
obtained by analytic continuation, 
\begin{eqnarray}
\Phi_{+}\left(\frac{\omega}{\Delta}\right) & = & \Theta\left(\left|\omega\right|-2\Delta\right)\frac{\sqrt{\omega^{2}-4\Delta^{2}}}{\omega}\times\label{eq:Phit_plus}\\
 &  & \left(1+\varepsilon\Phi_{2}\left(\frac{\omega}{\Delta}\right)\right)+\mathcal{O}\left(\varepsilon^{2}\right),\nonumber 
\end{eqnarray}
where 
\begin{eqnarray}
\Phi_{2}\left(\frac{\omega}{\Delta}\right) & = & \frac{1}{2}\ln\left(\frac{2\Delta+\frac{2\left|\omega\right|}{\sqrt{\omega^{2}-4\Delta^{2}}}}{\sqrt{\omega^{2}-4\Delta^{2}}-\left|\omega\right|}\right)\nonumber \\
 &  & +2\frac{N+2}{N+8}\frac{\sqrt{\omega^{2}-4\Delta^{2}}}{\left|\omega\right|}\tanh^{-1}\frac{\left|\omega\right|}{\sqrt{\omega^{2}-4\Delta^{2}}}\nonumber \\
 &  & +\frac{i}{2\pi}\text{Li}_{2}\left(\frac{1}{2}+\frac{\left|\omega\right|}{2\sqrt{\omega^{2}-4\Delta^{2}}}\right)\nonumber \\
 &  & -\frac{i}{2\pi}\text{Li}_{2}\left(\frac{1}{2}-\frac{\left|\omega\right|}{2\sqrt{\omega^{2}-4\Delta^{2}}}\right).
\end{eqnarray}
$\Phi_{+}\left(\frac{\omega}{\Delta}\right)$ is depicted in Fig.\ \ref{fig:Im=00007Bphi=00007D symmetric}.
We find that $\Phi_{+}\left(\frac{\omega}{\Delta}\right)$ has a threshold
at $\omega=2\Delta$, the minimal energy required to excite a pair
of quasiparticles with mass $\Delta$. To first order in $\varepsilon$,
$\Phi_{+}\left(\frac{\omega}{\Delta}\right)$ does not have a resonance,
unlike results obtained in $D=2+1$ using Quantum Monte Carlo \cite{chen2013universal}.

We examine $\Phi_{+}\left(\frac{\omega}{\Delta}\right)$ near the
threshold, at $\omega=2\Delta+\delta\omega$ as $\delta\omega\rightarrow0^{+}$
. We find
\begin{eqnarray}
\Phi_{+}\left(\frac{\delta\omega+2\Delta}{\Delta}\right) & \sim & \Theta\left(\delta\omega\right)\sqrt{\frac{\delta\omega}{\Delta}}\left[1-\frac{\varepsilon}{2}\ln\frac{\delta\omega}{\Delta}\right]\label{eq:Im=00007BS=00005Bp=00005D=00007D}\\
 & = & \Theta\left(\delta\omega\right)\left(\frac{\delta\omega}{\Delta}\right)^{\left(1-\varepsilon\right)/2}+\mathcal{O}\left(\varepsilon^{2}\right),\nonumber 
\end{eqnarray}
which agrees with Eq.\ (\ref{eq:AplusDef}), hence justifying the
choice of $\mathcal{A}_{+}$ in Eq.\ (\ref{eq:Aplus}). This power
law matches the expected behavior based on the density of states available
for exciting two counterpropagating bosons with total energy of $\omega=2\Delta+\delta\omega$,
\[
\int\frac{d^{D-1}k}{\left(2\pi\right)^{D-1}}\delta\left(\omega-2\sqrt{k^{2}+\Delta^{2}}\right)\propto\delta\omega^{\left(D-3\right)/2}\Theta\left(\delta\omega\right).
\]
 
\section{The Ordered Phase\label{sec:The-Ordered-Phase}}

We write the expectation value (EV) of the order parameter in the
ordered phase as 
\begin{equation}
\left\langle \phi^{2}\right\rangle =\frac{m^{2}}{U},\label{eq:EV_ordered}
\end{equation}
where $m^{2}=-2r+\mathcal{O}\left(\varepsilon\right)$. We parametrize
the fluctuations around the EV as, 
\begin{equation}
\phi=\left(\frac{m}{\sqrt{U}}+\sigma,\vec{\pi}\right).\label{eq:phi Goldstone}
\end{equation}
In Eq. (\ref{eq:phi Goldstone}), the fields $\sigma$ and $\pi$
represent the longitudinal (Higgs) and transverse (Goldstone) excitations
relative to the ordering direction, correspondingly.

The partition function in the ordered phase is obtained by inserting
Eq.\ (\ref{eq:phi Goldstone}) into Eq.\ (\ref{eq:Zu=00005BJ=00005D,phi}),
\begin{equation}
\mathcal{Z}\begin{array}{c}
=\int\MD\sigma\MD\pi\exp\left(-S_{0}-S_{C}-S_{I}\right)\end{array}\label{eq:Z Goldstone}
\end{equation}
where
\begin{eqnarray}
S_{0} & = & \frac{1}{2}\int_{x}\left[\left(\partial_{\mu}\pi\right)^{2}+\left(\partial_{\mu}\sigma\right)^{2}+m^{2}\sigma^{2}\right],\\
S_{C} & = & \frac{m^{2}+2r}{4U}\int_{x}\left[U\pi^{2}+U\sigma^{2}+2m\sqrt{U}\sigma\right],\nonumber \\
S_{I} & = & \int_{x}\left[\frac{1}{2}m\sqrt{U}\sigma\pi^{2}+\frac{1}{3!}3m\sqrt{U}\sigma^{3}+\right.\nonumber \\
 &  & \left.\frac{1}{4!}3U\sigma^{4}+\frac{1}{8}U\left(\pi^{2}\right)^{2}+\frac{1}{4}U\pi^{2}\sigma^{2}\right].\nonumber 
\end{eqnarray}
The terms $S_{0}$ and $S_{I}$ are the harmonic and interacting parts
of the action, respectively. The resulting tree-level Green's functions
are,
\[
\begin{array}{ccl}
G_{\sigma\sigma}^{0}\left(p\right) & = & \frac{1}{p^{2}+m^{2}},\\
\begin{array}{c}
G_{\pi\pi}^{0}\left(p\right)\end{array} & = & \frac{1}{p^{2}}.
\end{array}
\]
The term $S_{C}$ contains the counterterms. In principle, three separate
counterterms are necessary, one each for the terms $\pi^{2},$$\sigma^{2}$,
and $\sigma$. However, to$\mathcal{O}\left(\varepsilon\right)$,
all three are fixed by the requirement $\left\langle \sigma\right\rangle =0$.
At this order, this single condition guarantees that the Goldstone
modes are gapless and that the tree-level mass of the $\sigma$ field
is $m$. 

The condition $\left\langle \sigma\right\rangle =0$ yields, at $U_{c}$
\begin{eqnarray}
m^{2} & = & -2r-U_{c}\left(N-1\right)\int_{p}\frac{1}{p^{2}}-3U_{c}\int_{p}\frac{1}{p^{2}+m^{2}}+\mathcal{O}\left(\varepsilon^{2}\right)\nonumber \\
 & = & -2\delta r+\frac{3m^{2}}{2\left(N+8\right)}\varepsilon\ln\frac{\Lambda^{2}}{m^{2}}+\mathcal{O}\left(\varepsilon^{2}\right).\label{eq:m^2=00003Ddr}
\end{eqnarray}
In Eq.\ (\ref{eq:m^2=00003Ddr}), we have used the values of $r_{c}$,
see Eqs.\ (\ref{eq:Gff(p=00003D0)}). $m$ can be related to the
value of the gap in the partner point in the symmetric phase through
Eq.\ (\ref{eq:Symmetric phase pole}) as
\begin{equation}
m^{2}=2\Delta^{2}\left(1+\frac{\varepsilon}{2}\ln\frac{\Lambda^{2}}{2\Delta^{2}}+\frac{\varepsilon}{2}\frac{N+2}{N+8}\ln2\right)+\mathcal{O}\left(\varepsilon^{2}\right).\label{eq:Sigma0=00005BDelta=00005D}
\end{equation}
This indicates that $m^{2}$and $\Delta^{2}$ scale with different
exponents. This is expected. Despite the fact that $m^{2}$has units
of mass squared, Eq.\ (\ref{eq:EV_ordered}) shows that it scales
with the order parameter exponent $\beta$, rather than $\nu$.

\subsection{Scalar Susceptibility}

\begin{figure}[t]
\includegraphics[scale=0.47]{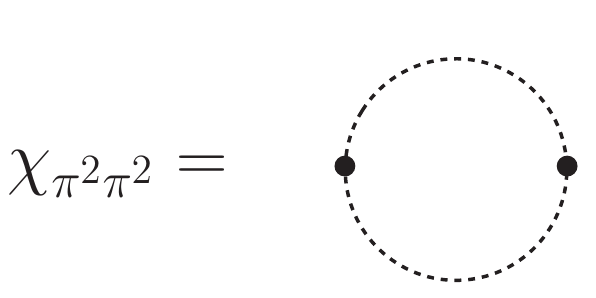} \quad{}\quad{}\enskip{}\quad{}\includegraphics[scale=0.47]{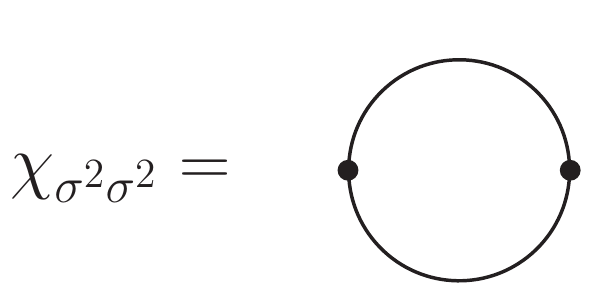}

\includegraphics[scale=0.47]{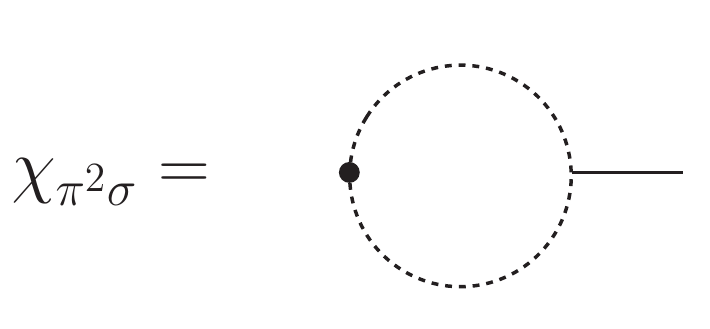}\quad{}\quad{}\includegraphics[scale=0.47]{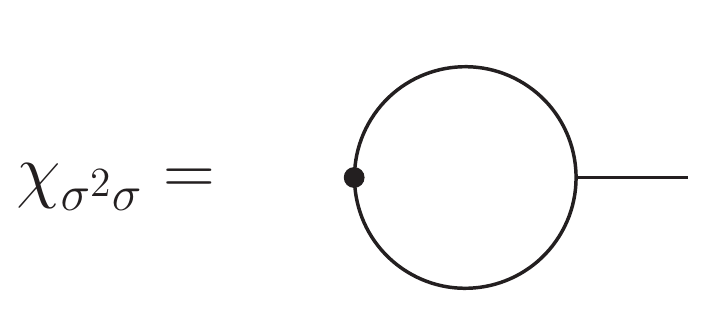}

$\begin{array}{c}
\chi_{\sigma\sigma}=\\
\\
\\
\\
\end{array}$\includegraphics[scale=0.52]{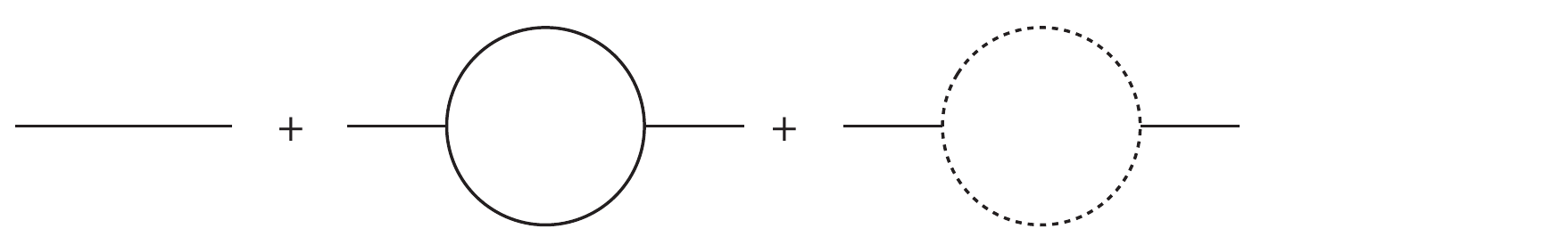}

\protect\caption{\label{fig:Diagramtic-expanssion-of-1}Diagrammatic expansion of the
scalar susceptibility $\chi_{s}$ to $\mathcal{O}\left(\varepsilon^{0}\right)$,
written in terms of the susceptibilities that compose it in the ordered
phase. $\chi_{\pi^{2}\pi^{2}}$ and $\chi_{\sigma^{2}\sigma^{2}}$
and are calculated to $\mathcal{O}\left(\varepsilon^{0}\right)$,
$\chi_{\sigma^{2}\sigma}$ and $\chi_{\pi^{2}\sigma}$ to $\mathcal{O}\left(\varepsilon^{1/2}\right)$
and $\chi_{\sigma\sigma}$ to $\mathcal{O}\left(\varepsilon\right).$
At this order, $\chi_{\pi^{2}\sigma^{2}}=0$.}
\end{figure}

In order to compute the scalar susceptibility in the ordered phase,
we use Eq.\ (\ref{eq:phi Goldstone}) to obtain, at $U_{c},$

\begin{equation}
\phi_{\alpha}^{2}\left(x\right)=\pi^{2}+\sigma^{2}+2\sigma\frac{m}{\sqrt{U_{c}}}+\frac{m^{2}}{U_{c}}.\label{eq:ro}
\end{equation}
Inserting this into Eq.\ (\ref{eq:S(P) definition}) yields,
\begin{eqnarray}
\begin{array}{c}
\chi_{s}\left(p\right)\end{array} & = & \chi_{\pi^{2}\pi^{2}}+\chi_{\sigma^{2}\sigma^{2}}+2\chi_{\pi^{2}\sigma^{2}}\label{eq:Chi_ro_ro}\\
 &  & +4\frac{m}{\sqrt{U_{c}}}\left(\chi_{\pi^{2}\sigma}+\chi_{\sigma^{2}\sigma}\right)+4\frac{m^{2}}{U_{c}}\chi_{\sigma\sigma}.\nonumber 
\end{eqnarray}
We evaluate $\chi_{s}\left(p\right)$ by summing over the different
susceptibilities in Eq.\ (\ref{eq:Chi_ro_ro}). Since $U_{c}=\mathcal{O}(\varepsilon)$,
the leading term in $\chi_{s}$ is of $\mathcal{O}\left(\varepsilon^{-1}\right)$.
Here we compute the scalar susceptibility to the next-to-leading order,
$\mathcal{O}\left(\varepsilon^{0}\right)$. This yields the diagrams
shown in Fig.\  \ref{fig:Diagramtic-expanssion-of-1}, which evaluate
to, 
\begin{eqnarray}
\Omega_{\varepsilon}^{-1}\chi_{s} & = & \frac{1}{\varepsilon}\frac{2m^{2}\left(N+8\right)}{p^{2}+m^{2}}+\frac{\left(N-1\right)p^{4}}{\left(p^{2}+m^{2}\right)^{2}}\left(1+\ln\frac{\Lambda^{2}}{p^{2}}\right)\label{eq:S=00005Bp=00005D ordered}\\
 &  & +\frac{\left(p^{2}-2m^{2}\right)^{2}}{\left(p^{2}+m^{2}\right)^{2}}\left(1+\ln\frac{\Lambda^{2}}{m^{2}}-\frac{2\tanh^{-1}x_{m}}{x_{m}}\right),\nonumber 
\end{eqnarray}
where $x_{m}=\frac{p}{\sqrt{p^{2}+4m^{2}}}$ . The calculation of
$\chi_{s}\left(p\right)$ to $\mathcal{O}\left(\varepsilon\right)$
is outlined in App.\ \ref{sec:Suscpetibilities Goldstone}.

\subsection{Universal Scaling Function \label{sub:Universal-Scaling-Function Ordered}}

Near the phase transition, Eq.\ (\ref{eq:S=00005Bp=00005D ordered})
can be written in the form
\[
\chi_{s}\left(p\right)=\mathcal{A}_{-}\Delta^{2\alpha}\tilde{\Phi}_{-}\left(\frac{p}{\Delta}\right)+\chi_{reg}
\]
where $\tilde{\Phi}_{-}\left(\frac{p}{\Delta}\right)$ is a universal
scaling function and $\chi_{reg}$ is the regular part. Since $\chi_{reg}$
is analytic across the transition, we can use the value of $\chi_{reg}$
obtained in the disordered phase, Eq.\ (\ref{eq:chireg}), as this
allows us to identify the universal part unambiguously. This simplifies
our analysis considerably.

Alternatively, one can repeat the analysis outlined in Sec.\ \ref{sec:The-Symmetric-Phase}
to extract the universal part and the regular parts, without previous
knowledge of $\chi_{reg}$. Then, in order to extract the universal
function unambiguously at $\mathcal{O}\left(\varepsilon^{0}\right)$
one must determine the logarithmic UV divergences to $\mathcal{O}\left(\varepsilon^{1}\right)$.
This is a difficult calculation (see App. \ref{sec:Suscpetibilities Goldstone}).
We have carried out this procedure, and found that indeed the value
of $\chi_{reg}$ obtained in this manner matches the disordered phase,
providing a very valuable consistency check of our calculations.

In order to obtain the universal scaling function, we use the known
value of $\chi_{reg}$ to rewrite Eq.\ (\ref{eq:S=00005Bp=00005D ordered})
in the form

\begin{eqnarray}
\Omega_{\varepsilon}^{-1}\chi_{s} & = & -\left[\frac{4}{\alpha_{1}}+\frac{2\left(N+8\right)p^{2}}{p^{2}+m^{2}}\right]\left[\frac{1}{\varepsilon}-\alpha_{1}\left(1+\ln\frac{\Lambda^{2}}{m^{2}}\right)\right]\nonumber \\
 &  & +N\frac{\alpha_{2}}{\alpha_{1}^{2}}+\frac{\left(N-1\right)p^{4}}{\left(p^{2}+m^{2}\right)^{2}}\left(1+\ln\frac{m^{2}}{p^{2}}\right)\nonumber \\
\begin{array}{c}
\end{array} &  & -\frac{2\left(p^{2}-2m^{2}\right)^{2}}{\left(p^{2}+m^{2}\right)^{2}}\frac{\tanh^{-1}x_{m}}{x_{m}}\label{eq:S(p) ordered 2}\\
 &  & -\frac{m^{2}p^{2}\left(N+8\right)}{\left(p^{2}+m^{2}\right)^{2}}\left(1+\ln\frac{\Lambda^{2}}{m^{2}}\right)+\Omega_{\varepsilon}^{-1}\chi_{reg},\nonumber 
\end{eqnarray}
where $\alpha_{1}$ is given in Eq.\ (\ref{eq:alpha1,2}) and $\chi_{reg}$
is given by Eq.\ (\ref{eq:chireg}). In the following step, we use
Eq.\ (\ref{eq:Sigma0=00005BDelta=00005D}) to eliminate $m$ in favor
of $\Delta$ in Eq.\ (\ref{eq:S=00005Bp=00005D ordered}),
\begin{eqnarray}
\Omega_{\varepsilon}^{-1}\chi_{s} & = & -\left[\frac{4}{\alpha_{1}}+\frac{2\left(N+8\right)p^{2}}{p^{2}+2\Delta^{2}}\right]\left[\frac{1}{\varepsilon}-\alpha_{1}\left(1+\ln\frac{\Lambda^{2}}{2\Delta^{2}}\right)\right]\nonumber \\
 &  & +N\frac{\alpha_{2}}{\alpha_{1}^{2}}-2\Delta^{2}p^{2}\frac{\left(N+2\right)\ln2-\left(N+8\right)}{\left(p^{2}+2\Delta^{2}\right)^{2}}\nonumber \\
 &  & +\frac{\left(N-1\right)p^{4}}{\left(p^{2}+2\Delta^{2}\right)^{2}}\left(1+\ln\frac{2\Delta^{2}}{p^{2}}\right)\nonumber \\
\begin{array}{c}
\end{array} &  & -\frac{2\left(p^{2}-4\Delta^{2}\right)^{2}}{\left(p^{2}+2\Delta^{2}\right)^{2}}\frac{\tanh^{-1}\tilde{x}}{\tilde{x}}+\Omega_{\varepsilon}^{-1}\chi_{reg},\label{eq:S(p) ordered delta}
\end{eqnarray}
where $\tilde{x}=\frac{p}{\sqrt{p^{2}+8\Delta^{2}}}$.

We obtain the overall constant $\mathcal{A}_{-}$ , 

\begin{equation}
\mathcal{A}_{-}=\Omega_{\varepsilon}\Lambda^{-2\alpha}\left(N-1\right)\frac{\pi}{4}\:.\label{eq:Aminus}
\end{equation}
which will be shown to agree with Eq.\ (\ref{eq:AminusDef}). Then, 

\begin{eqnarray}
\tilde{\Phi}_{-}\left(\frac{p}{\Delta}\right) & = & -\frac{2^{\alpha}}{\varepsilon}\frac{4\mathcal{C}}{N-1}\left(4+\frac{\left(N-4\right)p^{2}}{p^{2}+2\Delta^{2}}\right)\nonumber \\
 &  & +\frac{4}{\pi}\frac{p^{4}}{\left(p^{2}+2\Delta^{2}\right){}^{2}}\left(1+\ln\frac{2\Delta^{2}}{p^{2}}\right)\nonumber \\
 &  & +\frac{1}{\left(N-1\right)\pi}\frac{8\Delta^{2}p^{2}}{\left(p^{2}+2\Delta^{2}\right)^{2}}\left(-1+\frac{N+2}{N+8}\ln2\right)\nonumber \\
 &  & -\frac{8}{\left(N-1\right)\pi}\frac{\left(p^{2}-4\Delta^{2}\right){}^{2}}{\left(p^{2}+2\Delta^{2}\right){}^{2}}\frac{\tanh^{-1}\tilde{x}}{\tilde{x}}.\label{eq:eq:phi tilde ordered}
\end{eqnarray}
where $\mathcal{C}=\frac{1}{\alpha_{1}\pi}\left(\frac{1}{\varepsilon}-\alpha_{1}-\frac{\alpha_{2}}{\alpha_{1}}\right)$,
as before. The constant term $-\frac{2^{\alpha}}{\varepsilon}\frac{16\mathcal{C}}{N-1}$
in $\tilde{\Phi}_{-}\left(\frac{p}{\Delta}\right)$ does not contribute
to the imaginary part of $\tilde{\Phi}_{-}\left(-i\frac{\omega}{\Delta}+0^{+}\right)$
and will be omitted below.

As a check of our results, we consider the $p\gg\Delta$ limit, where
Eq.\ (\ref{eq:hyper_scale_limit}) is expected to hold. Indeed, we
find that
\begin{eqnarray}
\tilde{\Phi}_{-}\left(\frac{p}{\Delta}\right) & \to & \frac{4N}{N-1}\left(-2^{\alpha}\mathcal{C}+\frac{1}{\pi}\ln\frac{p^{2}}{2\Delta^{2}}\right)\label{eq:Chis_ordered_large_p}\\
 &  & +\mathcal{O}\left(\frac{\Delta}{p}\right)+\mathcal{O}\left(\varepsilon\right)\nonumber \\
 & = & -\frac{4N\mathcal{C}}{N-1}\left(\frac{p}{\Delta}\right){}^{2\alpha}+\mathcal{O}\left(\frac{\Delta}{p}\right)+\mathcal{O}\left(\varepsilon\right),\nonumber 
\end{eqnarray}
as expected to $\mathcal{O}(\varepsilon)$. Conversely, we consider
the low energy limit $\omega\ll\Delta$. In this limit, the universal
scaling function is expected to follow an asymptotic power-law behavior,
$\Phi\left(\frac{\omega}{\Delta}\right)\propto\left(\frac{\omega}{\Delta}\right){}^{4-\varepsilon}$,
due to the production of pairs of Goldstone modes\cite{podolsky2011visibility}.
Indeed, by writing $p=-i\omega+0^{+}$ in Eq.\ (\ref{eq:eq:phi tilde ordered})
and taking the limit $\omega\ll\Delta$, we find that the imaginary
part becomes

\begin{eqnarray*}
\Phi_{-}\left(\frac{\omega}{\Delta}\right) & \to & \left(\frac{\omega}{\Delta}\right)^{4-\varepsilon}+\mathcal{O}(\varepsilon^{1})
\end{eqnarray*}
as expected to this order in $\varepsilon$. This justifies our choice
of $\mathcal{A}_{-}$ in Eq.\ (\ref{eq:Aminus}), in agreement with
the definition in Eq.\  ($\ref{eq:AminusDef}$). These are non-trivial
consistency checks of our results.

The spectral function extracted directly from Eq.\ (\ref{eq:eq:phi tilde ordered})
diverges as $\omega\rightarrow\sqrt{2}\Delta$ in a non-integrable
manner. This can be avoided by noting that Eq.\ (\ref{eq:eq:phi tilde ordered})
has the form of a Dyson expansion to $\mathcal{O}\left(\varepsilon^{2}\right)$,
\begin{eqnarray*}
\frac{1}{p^{2}+2\Delta^{2}-\Sigma} & = & \frac{1}{p^{2}+2\Delta^{2}}+\frac{\Sigma}{\left(p^{2}+2\Delta^{2}\right)^{2}}\\
 &  & +\mathcal{O}\left(\varepsilon^{2}\right).
\end{eqnarray*}
Within $\mathcal{O}\left(\varepsilon^{0}\right)$, we can re-write

\begin{eqnarray}
\tilde{\Phi}_{-}\left(\frac{p}{\Delta}\right) & = & \frac{-2^{\alpha+2}\mathcal{C}\frac{N-4}{N-1}p^{2}}{p^{2}+2\Delta^{2}-\Sigma_{s}\left(p\right)},\label{eq:phi tilde2 ordered}
\end{eqnarray}
where
\begin{eqnarray}
\Sigma_{s}\left(p\right) & =- & \varepsilon\frac{p^{2}}{2}\frac{N-1}{N+8}\left(1+\ln\frac{2\Delta^{2}}{p^{2}}\right)\label{eq:sigmaS}\\
 &  & -\varepsilon\Delta^{2}\left(\frac{N+2}{N+8}\ln2-1\right)\nonumber \\
 &  & +\varepsilon\frac{\left(p^{2}-4\Delta^{2}\right)^{2}}{p^{2}\left(N+8\right)}\frac{\tanh^{-1}x_{\Delta}}{x_{\Delta}}.\nonumber 
\end{eqnarray}
In this form, Eq.\ (\ref{eq:phi tilde2 ordered}) has a well behaved
spectral function. The universal scaling function, $\Phi_{-}\left(\frac{\omega}{\Delta}\right)$,
is then extracted from $\tilde{\Phi}_{-}\left(\frac{p}{\Delta}\right)$
through the relation $\Phi_{-}\left(\frac{\omega}{\Delta}\right)=\Im\left\{ \tilde{\Phi}_{-}\left(-\frac{i\omega}{\Delta}\right)\right\} $.

\begin{figure}[h]
\includegraphics[width=1\columnwidth]{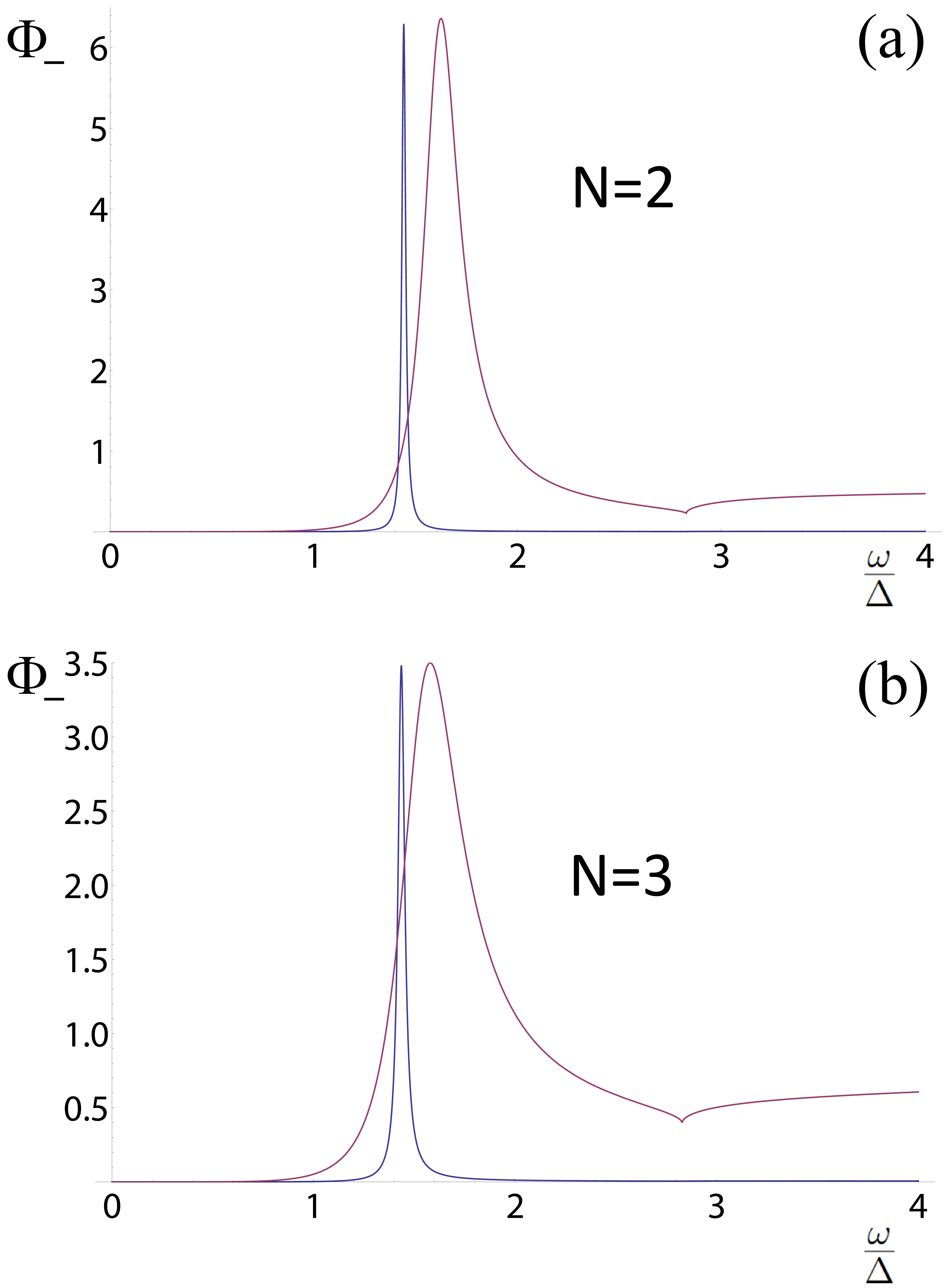}

\protect\caption{The universal scaling function in the ordered phase, $\Phi_{-}$$\left(-i\frac{\omega}{\Delta}+0^{+}\right)$
to first order in $\varepsilon$, divided by $2^{\alpha+2}\mathcal{C}\frac{N-4}{N-1}$.
Results are for $\varepsilon=0.1$ and $\varepsilon=1$, for $N=2$
(panel (a)) and $N=3$ (panel (b)) . $\Phi_{-}$ has a distinct peak
which corresponds to the Higgs excitation. As $\varepsilon$ is increased,
the peak broadens and its position is shifted towards higher frequencies.
At the threshold $\omega=2\sqrt{2}\Delta$, there exists a small kink
which is expected to be smoothened out at higher orders in $\varepsilon$.\label{fig:Im=00007Bphi=00007D ordered}}
\end{figure}

$\Phi_{-}\left(\frac{\omega}{\Delta}\right)$ is depicted in Fig.\ 
\ref{fig:Im=00007Bphi=00007D ordered} for $N=2$ and 3. $\Phi_{-}\left(\frac{\omega}{\Delta}\right)$
has a distinct peak which can be identified with the Higgs excitation.
The position of the peak depends on $\varepsilon$. For $\varepsilon=0$,
which is the noninteracting limit, the peak is a delta function at
the mean field value of the Higgs mass, $m_{H}/\Delta=\sqrt{2}$.
As $\varepsilon$ increases, the peak broadens and its position is
shifted towards larger frequencies. At $\varepsilon=1$, the Higgs
peak occurs at $m_{H}/\Delta=1.67$ $\left(N=2\right)$ and $m_{H}/\Delta=1.64$
$\left(N=3\right)$. 

\subsection{The Higgs pole}

\subsubsection{$D=4-\varepsilon$ for general $N$}

At $\varepsilon=0,$ the scaling function (\ref{eq:phi tilde2 ordered})
has a pole at $\omega=ip=\sqrt{2}\Delta$. At small $\varepsilon$,
the pole is shifted to the lower half complex plane, and a branch
cut starting at $p=0$ appears. In order to obtain this result, we
note that for small $\varepsilon$ the pole is expected to lie near
the real axis at $\omega_{{\rm pole}}=\sqrt{2}\Delta+\mathcal{O}\left(\varepsilon\right)$.
Substituting this expression into the denominator of Eq.\ (\ref{eq:phi tilde2 ordered})
and expanding to linear order in $\varepsilon$ yields

\begin{eqnarray}
\frac{\omega_{{\rm pole}}}{\sqrt{2}\Delta} & = & 1+\varepsilon\left(\frac{\left(N+2\right)\ln2+3\sqrt{3}\pi}{4\left(N+8\right)}-\frac{1}{4}\right)\nonumber \\
 &  & -i\varepsilon\frac{\pi}{4}\frac{N-1}{N+8}+\mathcal{O}\left(\varepsilon^{2}\right)\label{eq:wpole_eps}
\end{eqnarray}
which, at small $\varepsilon$, has a dominant real component. At
$\varepsilon=1$, we find $\omega_{{\rm pole}}/\Delta=1.74-0.11i$
for $N=2$, whereas $\omega_{{\rm pole}}/\Delta=1.70-0.20i$ for $N=3$.
These values match the position and width of the resonances plotted
in Fig.\ \ref{fig:Im=00007Bphi=00007D ordered}, up to the corrections
of order $\varepsilon^{2}$ in Eq.\ (\ref{eq:wpole_eps}). Note that
these values are smaller than found in previous analyses. For example,
two separate QMC analyses found $m_{H}/\Delta=2.1(3)$ for $N=2$
and $m_{H}/\Delta=2.2(3)$ for $N=3$ \cite{gazit2013fateof,gazit2013dynamics};
and $3.3\left(8\right)$ for $N=2$ and $3.2\left(8\right)$ for $N=3$
\cite{chen2013universal}; whereas an NPRG analyses obtained $m_{H}/\Delta\approx2.5$
for $N=2$ \cite{RanconDupuis}. Qualitatively, the results agree
on the sign of the shift in the Higgs mass relative to the mean field
value $\sqrt{2}\Delta$, and on the fact that the Higgs mass is similar
for $N=2$ and $N=3$. Quantitatively, the comparison of Higgs masses
is not straightforward since higher order corrections to our results
may be significant at $\varepsilon=1$. We note that there is substantial
disagreement between the different numerical analysis, and that our
results are easier to reconcile with the lower of those results.

From the real and imaginary parts of Eq.\ (\ref{eq:wpole_eps}) we
can read off the Higgs mass $m_{{\rm H}}$ and its decay rate $\Gamma_{{\rm H}}$,
respectively. This yields the universal ratios:

\[
\frac{m_{{\rm H}}}{\sqrt{2}\Delta}=1+\varepsilon\left(\frac{\left(N+2\right)\ln2+3\sqrt{3}\pi}{4\left(N+8\right)}-\frac{1}{4}\right),
\]

\[
\frac{\Gamma{}_{{\rm H}}}{\sqrt{2}\Delta}=\varepsilon\frac{\pi}{4}\frac{N-1}{N+8}.
\]
At this order in $\varepsilon$, $m_{H}/\Delta$ is monotonically
decreasing with $N$ for all $\varepsilon>0,$ whereas it increases
with $\varepsilon$ for $N<32$. On the other hand, $\Gamma{}_{{\rm H}}/\Delta$
grows with $\varepsilon$, leading to a broadening of the resonance
with lower dimension, as expected from the increased coupling strength
as we move away from the Gaussian fixed point at $D=3+1$. In addition,
the width of the resonance grows with $N$, reflecting the larger
number of Goldstone modes into which the Higgs mode can decay. The
ratio $m_{{\rm H}}/\Gamma_{H}$ therefore grows monotonically with
$N$, but is found to saturate at large $N$. 

A previous NPRG calculation \cite{RanconDupuis} found that in two
spatial dimensions $\left(\varepsilon=1\right)$, the Higgs mode yields
a distinct peak for $N=2$, but that it is strongly suppressed for
$N\ge3$. By contrast, QMC simulations found a Higgs peak for both
$N=2$ and $N=3$ \cite{gazit2013dynamics,gazit2013fateof}. In our
analysis, we find that for $\varepsilon<\frac{4}{1+\pi-\ln2}\approx1.159$,
the ratio $m_{{\rm H}}/\Gamma_{H}$ is larger than one for all values
of $N$, indicating that the resonance exists for any value of $N$.
However, for $\varepsilon>1.159$, the condition $m_{H}/\Gamma_{H}>1$
is no longer satisfied for large values of $N$, and therefore is
possible that a calculation to higher orders in $\varepsilon$ could
change this picture.

A similar analytic structure was also found in a large $N$ analysis
in $D=2+1$ dimensions\cite{podolsky2012spectral}. There, it was
obtained that at $N\rightarrow\infty$, $\omega_{{\rm pole}}=-4i\Delta/\pi$
is purely imaginary at $D=2+1$. In the large $N$ expansion, the
pole moves away from the imaginary axis, where the deviation from
the imaginary axis is of $\mathcal{O}\left(\frac{1}{N}\right)$. We
will now study the evolution of $\omega_{{\rm pole}}$ with the dimension
of the system in the $N\rightarrow\infty$ limit.

\subsubsection{$N\to\infty$ for general $D$}

Following Ref.\cite{podolsky2012spectral} we can use a Hubbard-Stratonovich
transformation to write the action in the ordered phase, Eq.\ (\ref{eq:Action}),
in the form,

\begin{eqnarray}
\mathcal{Z} & = & \int\MD\sigma\MD\lambda\exp\left(-\mathcal{S}{}_{0}\right),
\end{eqnarray}
where
\begin{eqnarray}
\mathcal{S}{}_{0} & = & \int\frac{d^{D}p}{\left(2\pi\right)^{D}}\left[p^{2}\sigma^{2}+2i\sigma_{0}\sigma\lambda\right.\\
 &  & \left.+\frac{1}{2}\left(\Pi_{D}\left(p,0\right)+\frac{2}{NU}\right)\lambda^{2}+\mathcal{O}\left(\frac{1}{N}\right)\right].\nonumber 
\end{eqnarray}
Here, $\sigma_{0}$ is given by 
\begin{eqnarray}
\sigma_{0}^{2} & = & \int\frac{d^{D}p}{\left(2\pi\right)^{D}}\frac{1}{p^{2}}-\int\frac{d^{D}p}{\left(2\pi\right)^{D}}\frac{1}{p^{2}+r}\label{eq:mD}\\
 & = & -\frac{\pi}{2}\frac{K_{D}}{\sin\left(\frac{\pi D}{2}\right)}r^{-1+D/2}.\nonumber 
\end{eqnarray}
and 
\begin{equation}
\Pi_{D}\left(p,0\right)=\left(2-D\right)\frac{\pi}{4}\frac{p^{D-4}K_{D}\Gamma\left(\frac{D}{2}-1\right)^{2}}{\sin\left(\frac{\pi D}{2}\right)\Gamma\left(D-2\right)}\label{eq:PiD}
\end{equation}
is the massless polarization bubble in a general space-time dimension
$D$. Note that in this section, we use dimensional regularization
in order to regularize UV divergencies.

The scalar susceptibility is related to the two point function of
the $\lambda$ field\cite{podolsky2012spectral} by the equation
\[
\chi_{s}\left(p\right)=\frac{4N}{U}\left(1-\frac{N}{U}G_{\lambda\lambda}\left(p\right)\right),
\]
where the bare connected Green's function for the $\lambda$ field
is given by
\begin{equation}
G_{\lambda\lambda}\left(p\right)=\frac{2p^{2}}{\frac{p^{2}}{2}\left(\Pi_{D}\left(p,0\right)+\frac{2}{NU}\right)+2\sigma_{0}^{2}}.\label{eq:Gll}
\end{equation}
As a consistency check with our $D=4-\varepsilon$ calculation, we
find that for small $\varepsilon$ and in the large $N$ limit, Eqs.\ (\ref{eq:phi tilde2 ordered})
and (\ref{eq:Gll}) agree up to an overall constant. Following the
scheme which is outlined in Ref. \cite{podolsky2012spectral}, we
omit from Eq.\ (\ref{eq:Gll}) the term $\frac{2}{NU}$, which is
negligible compared to $\Pi_{D}\left(p,0\right)$ at small $p$.

\begin{figure}[h]
\includegraphics[width=1\columnwidth]{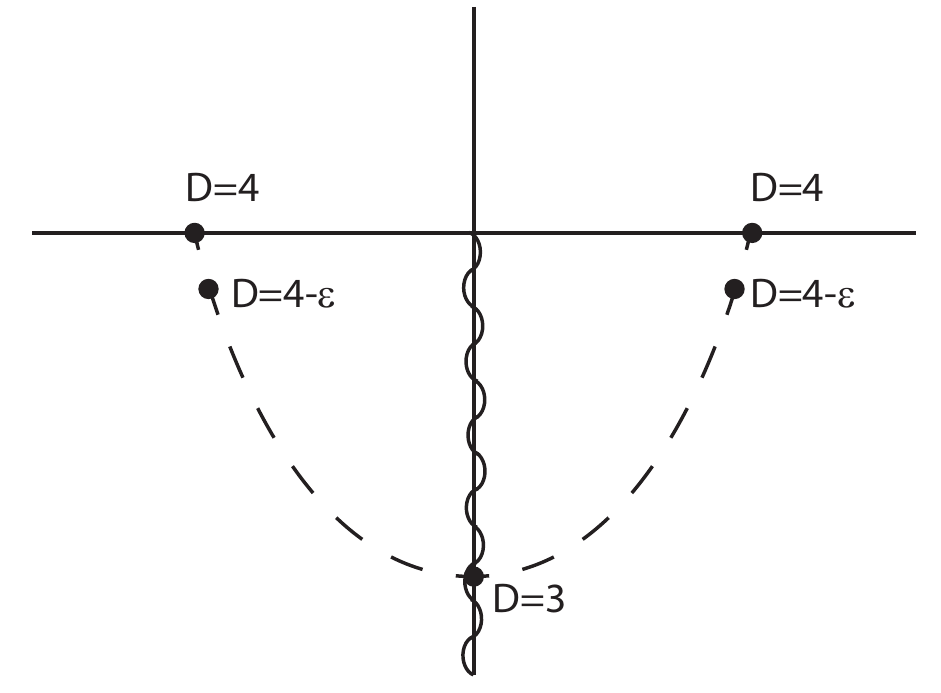}

\protect\caption{Analytic structure of the universal scaling function $\Phi_{-}(\omega)$
in the complex $\omega$ plane, in the $N\to\infty$ limit. In addition
to a branch cut starting at $\omega=0$, there are poles in the lower
half plane whose positions depend on the space time dimension $D$.
Note that the poles appear as partners on mirror positions on either
side of the branch cut. For $D=3+1$, the pole lies on the real axis,
at $\sqrt{2}\Delta$. For $D=2+1$ and $N=\infty$, $\omega_{{\rm pole}}=-4i\Delta/\pi$
is purely imaginary. For $D=2+1$ and large but finite $N$, the pole
is shifted away from the imaginary axis\cite{podolsky2012spectral}
by an amount of $\mathcal{O}\left(\frac{1}{N}\right)$. \label{fig:Pole_Large_N}}
\end{figure}

Let us now explore the evolution of the $N\rightarrow\infty$ pole
as we change the space-time dimension from $D=2+1$ to $D=3+1$. In
the $N\rightarrow\infty$ limit, $r_{c}=0$ and we can obtain the
equation for the quasi-particle pole by applying the relations 
\begin{equation}
\delta r=r=\Delta^{2}+\mathcal{O}\left(\frac{1}{N}\right)\label{eq:dr}
\end{equation}
together with Eq.\ (\ref{eq:mD}). We insert Eqs.\ (\ref{eq:mD}),
(\ref{eq:PiD}) and (\ref{eq:dr}) into Eq.\ (\ref{eq:Gll}) to obtain
that for a general space-time dimension $D$,
\begin{equation}
\frac{\omega_{{\rm pole}}}{2\Delta}=\left[\frac{i^{D}}{\pi^{3/2}}\Gamma\left(\frac{2-D}{2}\right)\Gamma\left(\frac{D-1}{2}\right)\sin\frac{\pi D}{2}\right]^{\frac{1}{D-2}}.\label{eq:Wpole}
\end{equation}
As expected, we find that at $D=2+1$, $\omega_{{\rm pole}}=-4i\Delta/\pi$
is purely imaginary. As $D$ is increased from $D=2+1$ to $D=3+1$,
$\omega_{{\rm pole}}$ acquires a real part which becomes more dominant
as $D$ approaches $4$, where $\omega_{{\rm pole}}=\sqrt{2}\Delta$
becomes purely real, see Fig.\ \ref{fig:Pole_Large_N}. This demonstrates
that the pole obtained in a $1/N$ expansion at $D=2+1$ in Ref.\cite{podolsky2012spectral}
can indeed be identified with the Higgs mode, as it smoothly evolves
with dimension to the sharply-defined Higgs resonance at $D=3+1$. 

\section{Quantum Critical Regime\label{sec:Expansion-in-Finite}}

\begin{figure}
\includegraphics[width=0.7\columnwidth]{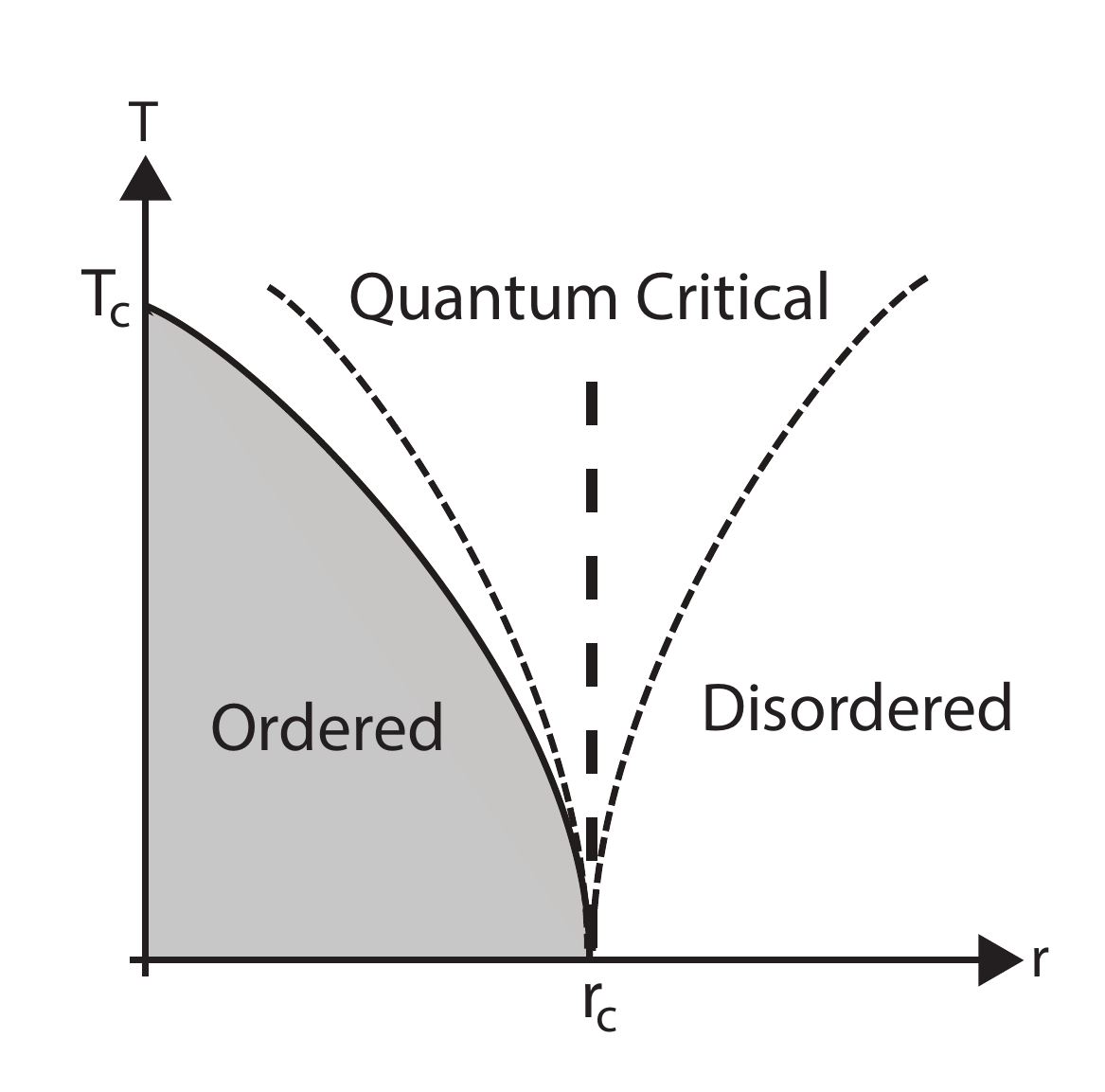}\protect\caption{The phase diagram of the $O\left(N\right)$ model in the $r-T$ plane.
We present results along the thick dashed line. \label{fig:Phase Diagram}}
\end{figure}

We extend the calculation of the scalar susceptibility to the quantum
critical regime, at temperature $T>0$ and along the line $\delta r=0$
(see Fig.\ \ref{fig:Phase Diagram}). Since the order parameter vanishes
in this regime, the formalism is similar to that given in Sec.\ \ref{sec:The-Symmetric-Phase}.
The calculations at $T>0$ are performed by discretizing the $q^{0}$
component of the Euclidean energy-momentum vector in Matsubara frequencies,
\[
\mathbf{q}=\left(\omega_{n},\vec{q}\right)
\]
with $\omega_{n}=2\pi nT$, where $n$ is an integer. The integral
$dq_{0}$ is then replaced by a sum,
\[
\int\frac{d^{4-\varepsilon}q}{\left(2\pi\right)^{4-\varepsilon}}\rightarrow T\sum_{n=-\infty}^{\infty}\int\frac{d^{3-\varepsilon}q}{\left(2\pi\right)^{3-\varepsilon}}.
\]
At $T>0$, thermal fluctuations shift the location of the critical
point to enlarge the disordered phase, see Fig.\ \ref{fig:Phase Diagram}.
The resulting correction to the self energy of the $\phi$ field is,
to first order,
\begin{eqnarray}
\Sigma & = & -\frac{U_{c}\left(N+2\right)}{2}\left[T\sum_{\omega_{n}}\int\frac{d^{3-\varepsilon}q}{\left(2\pi\right)^{3-\varepsilon}}\frac{1}{\omega_{n}^{2}+\vec{q}^{2}}\right.\label{eq:dr1}\\
 &  & \left.-\int\frac{d^{4-\varepsilon}q}{\left(2\pi\right)^{4-\varepsilon}}\frac{1}{q^{2}}\right]\nonumber \\
 & = & -\varepsilon\frac{N+2}{N+8}\frac{T^{2}}{12K_{4}}+\mathcal{O}\left(\varepsilon^{2}\right)\equiv-m_{T}^{2}.\nonumber 
\end{eqnarray}
The transition then occurs when $\delta r-\Sigma=0$, which corresponds
to $\delta r=-m_{T}^{2}$. Conversely, at $\delta r=0$ the $\phi$
propagator has an effective mass $m_{T}$, 
\[
G\left(\omega_{n},\vec{q};\delta r=0,T\right)=\frac{1}{\omega_{n}^{2}+\vec{q}^{2}+m_{T}^{2}}+\mathcal{O}\left(\varepsilon^{2}\right).
\]
Note that if higher order corrections are included, the self energy
is expected to become momentum dependent.

We next turn to evaluate the scalar susceptibility. The formal expression
for $\chi_{scalar}$ in the disordered phase was obtained in terms
of the polarization bubble $\Pi\left(p,\delta r\right)$ in Sec.\ \ref{sub:Scalar-Susceptibility}.
This result can be generalized to finite temperatures by replacing
$\Pi\left(p,\delta r\right)$ with $\Pi_{T}\left(\omega_{n},\vec{p},\delta r\right)$
in Eq.\ (\ref{eq:S(p) symmetric scehmatic}), 
\begin{eqnarray}
\chi_{s}\left(\omega_{n},\vec{p},T\right) & = & 2N\Pi_{T}\left(\omega_{n},\vec{p},\delta r\right)\label{eq:S(p) T>0}\\
 &  & -U_{c}N\left(N+2\right)\Pi_{T}\left(\omega_{n},\vec{p},\delta r\right)^{2}+\mathcal{O}\left(\varepsilon^{2}\right).\nonumber 
\end{eqnarray}
Here, $\Pi_{T}\left(\omega_{n},\vec{p},\delta r\right)$ is the polarization
bubble at finite temperatures. We focus on $\vec{p}=0$ and $\delta r=0$.
The formal expression for $\Pi_{T}^{0}(\omega_{n})\equiv\Pi_{T}\left(\omega_{n},\vec{p}=0,\delta r=0\right)$
is then

\begin{eqnarray}
\Pi_{T}^{0}\left(\omega_{n}\right) & = & \frac{T}{\left(2\pi\right)^{3}}\sum_{m}\int\frac{d^{3-\varepsilon}q}{\left(2\pi\right)^{3}}\frac{1}{\omega_{m}^{2}+\vec{q}^{2}+m_{T}^{2}}\label{eq:Pi General High_T}\\
 &  & \times\frac{1}{\left(\omega_{m}-\omega_{n}\right)^{2}+\vec{q}^{2}+m_{T}^{2}},\nonumber 
\end{eqnarray}

Temperature has the effect of regularizing IR divergences. However,
it has no effect on UV divergences, which must therefore be the same
for $\Pi_{T}^{0}\left(\omega_{n},\vec{p}\right)$ and $\Pi\left(p,\Delta\right)$,
Eq.\ (\ref{eq:pi(p,r)}). This implies that $\Pi_{T}^{0}\left(\omega_{n}\right)$
can be written in the form
\begin{eqnarray}
\Omega_{\varepsilon}^{-1}\Pi_{T}^{0}\left(\omega_{n}\right) & = & \pi_{0}\left(\omega_{n}\right)\left(1+\frac{\varepsilon}{2}\left(1+\ln\frac{\Lambda^{2}}{T^{2}}\right)\right)\nonumber \\
 &  & +\frac{1}{2}\ln\frac{\Lambda^{2}}{T^{2}}+\frac{\varepsilon}{8}\ln^{2}\frac{\Lambda^{2}}{T^{2}}\nonumber \\
 &  & +\varepsilon\pi_{1}\left(\omega_{n}\right)+\mathcal{O}\left(\varepsilon^{2}\right)\label{eq:P=00005Bp,T=00005D, geneal form}
\end{eqnarray}
where $\pi_{0}$ and $\pi_{1}$ are functions that are independent
of the cutoff and of $\varepsilon$, as we have verified explicitly.

We can now obtain the general form of the scalar susceptibility for
$T>0$ by inserting Eq.\ (\ref{eq:P=00005Bp,T=00005D, geneal form})
into Eq.\ (\ref{eq:S(p) symmetric scehmatic}), 
\begin{eqnarray}
\frac{\Omega_{\varepsilon}^{-1}}{N}\chi_{s}\left(\omega_{n},T\right) & = & \ln\frac{\Lambda^{2}}{T^{2}}-\frac{1}{2}\alpha_{1}\varepsilon\ln^{2}\frac{\Lambda^{2}}{T^{2}}\label{eq:S=00005Bp=00005D High_T}\\
 &  & +2\pi_{0}\left(1-\alpha_{1}\varepsilon\ln\frac{\Lambda^{2}}{T^{2}}\right)\nonumber \\
 &  & -\varepsilon\left(1+2\alpha_{1}\right)\pi_{0}^{2}\nonumber \\
 &  & +\varepsilon\left(2\pi_{1}+\pi_{0}\right)+\mathcal{O}\left(\varepsilon^{2}\right).\nonumber 
\end{eqnarray}

\begin{figure}
\includegraphics[width=1\columnwidth]{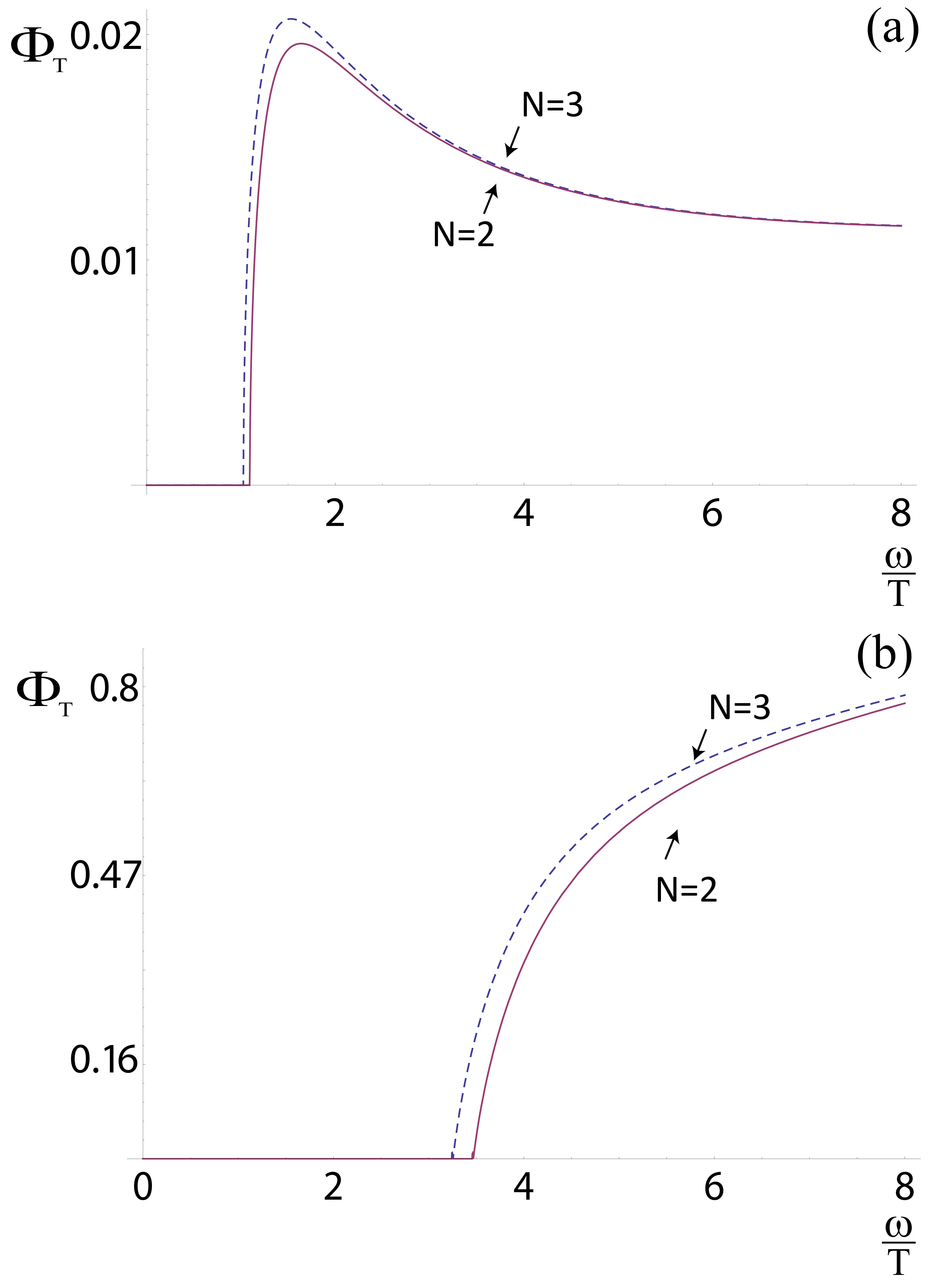}

\protect\caption{The universal scaling function in the quantum critical regime, \textbf{$\Phi_{T}\left(\frac{\omega}{T}\right)$},
to second order in $\varepsilon$, divided by $N$. Results are for
$\varepsilon=0.1$ (panel (a)), $\varepsilon=1$ (panel (b)) and for
$N=2,3$. \label{fig:Im=00007Bphi=00007D symmetric_highT}}
\end{figure}

We extract the universal scaling function from Eq.\ (\ref{eq:S=00005Bp=00005D High_T})
by proceeding in line with Sec.\ \ref{Appendix:Universal part Symmetric}.
As before, we obtain a logarithmic dependence on $\Lambda$ with prefactor
of order $\varepsilon^{0}$. This motivates us to choose $\mathcal{A}_{T}=\mathcal{A}_{+}$,
as given in Eq.\ (\ref{eq:Aplus}). We find that
\begin{eqnarray}
\Phi_{T}\left(\frac{\omega}{T}\right) & = & \frac{2}{\pi}\left(\pi_{0}^{''}\left(\omega\right)+\varepsilon\pi_{1}^{''}\left(\omega\right)\right)\label{eq:_Scaling_High_T}\\
 &  & -\frac{2}{\pi}\varepsilon\left(1+2\alpha_{1}\right)\pi_{0}^{'}\left(\omega\right)\pi_{0}^{''}\left(\omega\right)\nonumber \\
 &  & +\varepsilon\alpha_{1}+\frac{\varepsilon}{2}\ln2+\mathcal{O}\left(\varepsilon^{2}\right).\nonumber 
\end{eqnarray}
where $\pi_{0}^{'}=\Re\left\{ \pi_{0}\left(-i\omega+0^{+}\right)\right\} $$,\pi_{0}^{''}=\Im\left\{ \pi_{0}\left(-i\omega+0^{+}\right)\right\} $
and $\pi_{1}^{''}=\Im\left\{ \pi_{1}\left(-i\omega+0^{+}\right)\right\} $.
In order to obtain the explicit form of Eq.\ (\ref{eq:_Scaling_High_T}),
we are left to calculate $\pi_{0}^{'}$, $\pi_{0}^{''}$ and $\pi_{1}^{''}$.
The calculation of the latter two is given in Appendix \ref{appendix:COMPUTATION-OF-},
while $\pi_{0}^{'}$ is obtained numerically. We find 
\begin{eqnarray}
\pi_{0}^{''}\left(\omega\right) & = & \Theta\left(\left|\omega\right|-2m_{T}\right)\frac{\pi\sqrt{\omega^{2}-4m_{T}^{2}}}{2\left|\omega\right|}\coth\frac{\omega}{4T},\label{eq:Impi0}
\end{eqnarray}
\begin{eqnarray}
\pi_{1}^{''}\left(\omega\right) & = & -\pi_{0}^{''}\left(\omega\right)\times\frac{1}{2}\ln\left|\frac{\omega^{2}}{4}-m_{T}^{2}\right|.\label{eq:Impi0-1}
\end{eqnarray}
We can now obtain the explicit form of the universal scaling function
for $T>0$, to second order in $\varepsilon$, by inserting Eqs.\ (\ref{eq:Impi0})
and (\ref{eq:Impi0-1}) into Eq.\ (\ref{eq:_Scaling_High_T}),
\begin{eqnarray}
\Phi_{T}\left(\frac{\omega}{T}\right) & = & \Theta\left(\left|\omega\right|-2m_{T}\right)\frac{\sqrt{\omega^{2}-4m_{T}^{2}}}{\left|\omega\right|}\coth\frac{\omega}{4T}\nonumber \\
 &  & \times\left[1-2\varepsilon\left(1+2\alpha_{1}\right)\pi_{0}^{'}\left(\omega\right)\right.\nonumber \\
 &  & \left.+\frac{\varepsilon}{2}\ln\left|\frac{\omega^{2}}{4}-m_{T}^{2}\right|\right]+\mathcal{O}\left(\varepsilon^{2}\right).\label{eq:Phi_T}
\end{eqnarray}
As a consistency check, we find that in the high frequency limit,
$\omega\gg T$, Eq.\ (\ref{eq:Phi_T}) yields the expected scaling
form, $\Phi_{T}\propto(\omega/T)^{2\alpha}$.

Equation (\ref{eq:Phi_T}) is plotted in Fig.\ \ref{fig:Im=00007Bphi=00007D symmetric_highT}.
For $\varepsilon=0.1$, the response function $\Phi_{T}\left(\frac{\omega}{T}\right)$
exhibits a Higgs-like peak near the threshold. However, this peak
broadens as $\varepsilon$ is increased and is no longer present at
$\varepsilon=1$.

Note that $\Phi_{T}\left(\frac{\omega}{T}\right)$ has a threshold
at $\omega=2m_{T}$. Unlike the threshold at $T=0$ in the disordered
phase, which is a consequence of the gap in the spectrum in that case,
in the quantum critical regime the spectrum is gapless and hence no
such threshold is expected. In fact, the threshold is an artifact
of our working order, in which the mass term $m_{T}$ is independent
of momentum, hence playing the role of a hard gap. At two loop level,
the self energy becomes momentum dependent, hence smearing the threshold.
This calculation is difficult and it may not be possible to extract
the low frequency response function for $\omega<\sqrt{\varepsilon}T$
reliably from such a calculation \cite{subirCrossover}. However,
QMC simulations for $N=3$ indicate that the quasiparticle width in
the quantum critical region is small, and hence a threshold-like effect
may still exist even after higher order corrections are taken into
account \cite{PhysRevB.61.3475}.

\section{Summary and Discussion\label{sec:Summary-and-Conclusions}}

We have studied the Higgs mode of the relativistic $O(N)$ model near
$D=3+1$ spacetime dimensions by computing the scalar spectral function
through a controlled expansion in the small parameter $\varepsilon=4-D$,
and extracting the universal scaling function near the quantum phase
transition between the ordered and disordered phases.

In the ordered phase, the spectral function has a complex pole associated
with a sharp Higgs resonance. The pole occurs at a strictly real frequency
at $D=3+1$, and acquires a small imaginary component for $D=4-\varepsilon$.
Previous work computed the scalar susceptibility in the large $N$
limit and found a pole corresponding to a damped excitation at $D=2+1$,
which was identified with the Higgs resonance \cite{podolsky2012spectral}.
Indeed, we show that for $N\to\infty,$ this pole evolves smoothly
with $D$ to the sharp Higgs mode seen at $D=4-\varepsilon$.

Our analysis predicts a sharp Higgs mode in three spatial dimensions
close to the critical point. However, this does not indicate that
the Higgs mode has zero width in realistic experiments \cite{ruegg2008quantum,PhysRevB.84.134418}.
While our calculation applies asymptotically close to the QCP, for
$D=3+1$ the relative width of the Higgs mode approaches zero only
logarithmically in $\delta r$, and hence even relatively close to
the QCP, the Higgs resonance has a finite width\cite{PhysRevB.84.134418,affleck1992longitudinal}.

In the disordered phase, we have found that the scalar spectral function
has a threshold at $\omega=2\Delta$ and no Higgs-like peak. This
outcome is in disagreement with a previous QMC analysis which found
a peak close to the threshold, which was interpreted as a precursor
to the Higgs mode in the disordered phase \cite{chen2013universal}.
A weak peak in the spectral function was also found in a separate
QMC analysis \cite{gazit2013dynamics}, as well as in a NPRG calculation
\cite{RanconDupuis}. However, the spectral weight of this peak is
much smaller than that of the Higgs peak on the ordered side. We note,
furthermore, that these analyses rely on numerical analytic continuation,
which is difficult to control and which is liable to yield spurious
oscillations when the spectral function changes rapidly \cite{Beach}.
Of course, our calculation is only controlled for small $\varepsilon$,
and it cannot conclusively rule out such a peak for $D=2+1$. Finally,
we note that even if such a peak were to be present, renormalization
group arguments show that it should not be interpreted as a precursor
to the Higgs mode \cite{RanconDupuis}. Instead, such a peak could
be an indication of an emergent a bound state of gapped particle-hole
excitations in the disordered phase near the QCP.

Finally, we computed the universal spectral function in the quantum
critical regime. For $\varepsilon\ll1$, we find indirect evidence
for a peak in the spectral function at $\omega$ of order $T$, which
may agree with Ref. \cite{chen2013universal}, although for $\varepsilon=1$
no such peak is seen. Instead, only a threshold-like behavior is observed
at low frequencies.

It would be interesting to apply these methods to study other dynamical
properties such as the reactive conductivity\cite{gazit2014capacitance}
near quantum criticality in $D=4-\varepsilon$ dimensions.

\section*{Acknowledgments}

We would like to thank Shmuel Fishman, Snir Gazit, Adam Rancon, Subir
Sachdev, and William Witczak-Krempa for useful discussions. D. P.
acknowledges support from an Israeli Science Foundation grant and
from a joint grant of the Israel Science Foundation and the Indian
University Grant Commission. We thank the Aspen Center for Physics,
where part of this work was completed.

\vspace{3cm}

\onecolumngrid

\appendix

\section{Scalar Susceptibility in the Ordered Phase \label{sec:Suscpetibilities Goldstone}}

\begin{figure}[h]
\includegraphics[width=0.55\columnwidth]{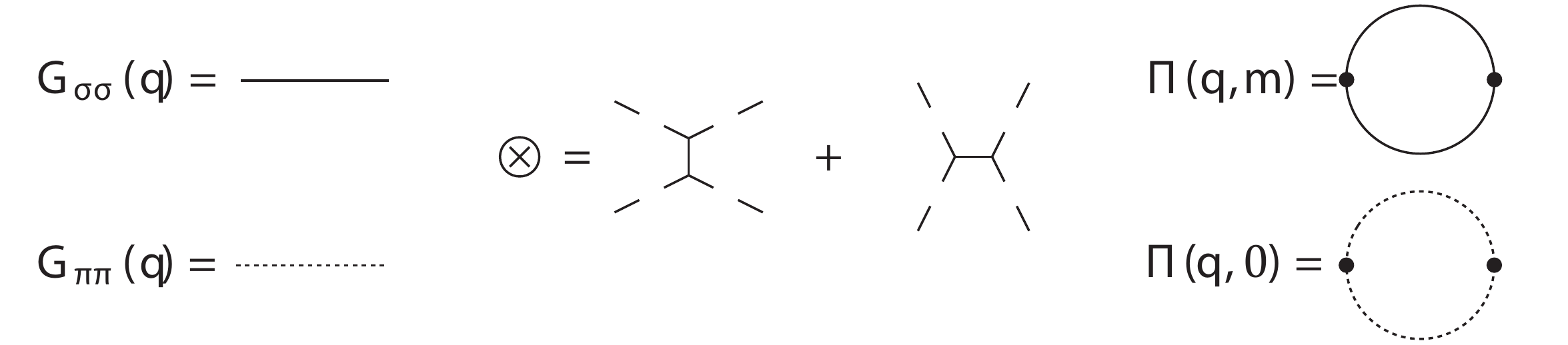}

\protect\caption{Notations for the Feynman Diagrams in the ordered phase. The cross
represents the different ways to contract lines at the interaction
vertex.\label{fig:The-standard-notations}}
\end{figure}

\begin{figure}[h]
\includegraphics[scale=0.47]{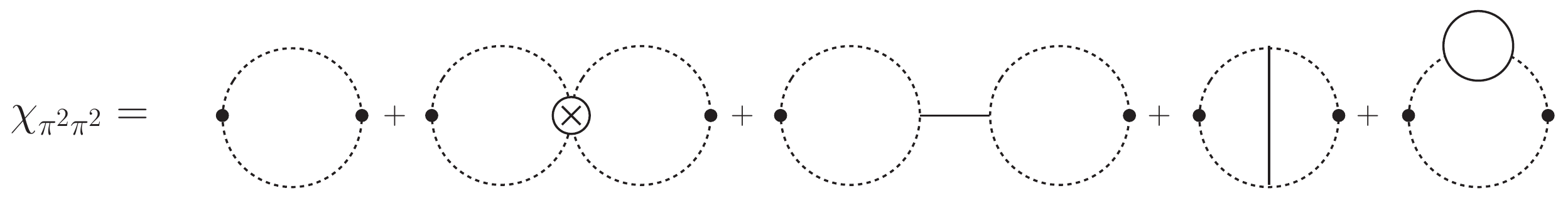}

\includegraphics[scale=0.47]{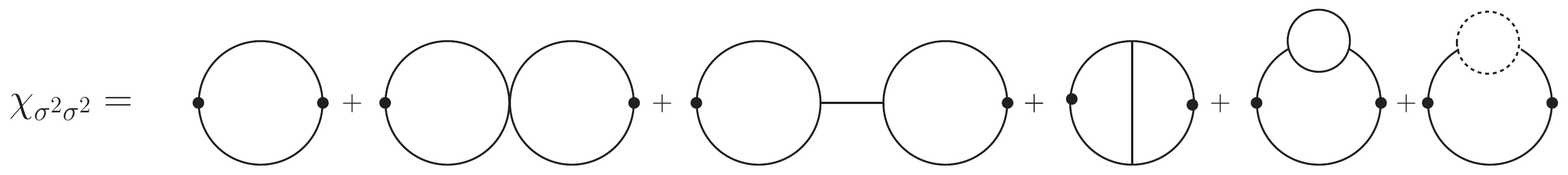}

\includegraphics[scale=0.47]{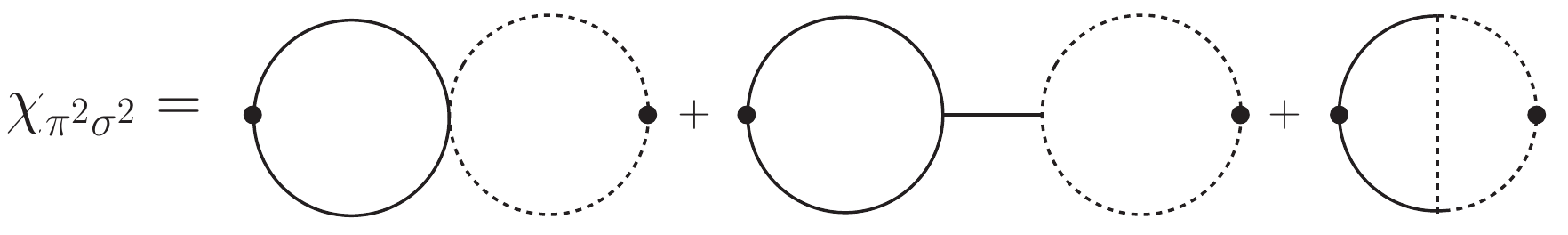}

\includegraphics[scale=0.47]{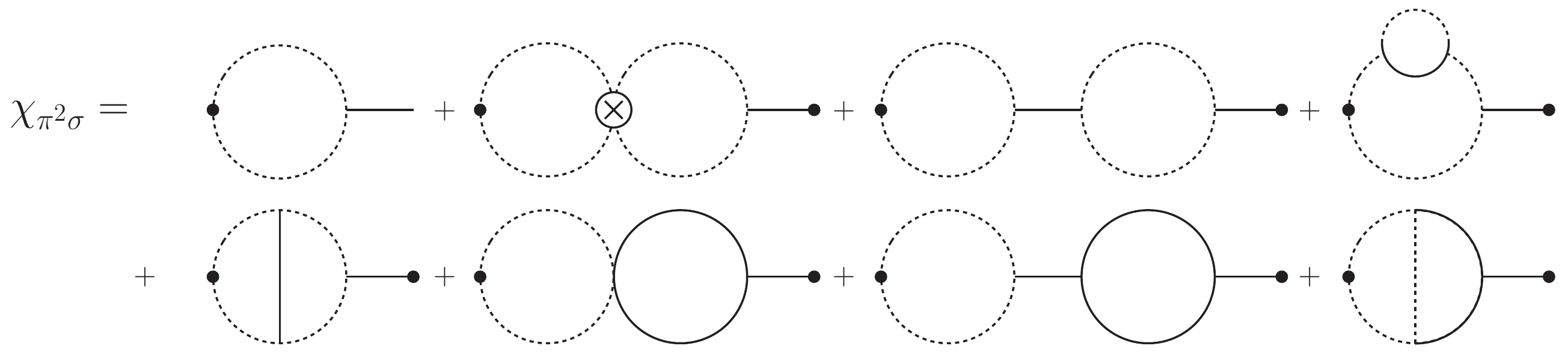}

\includegraphics[scale=0.47]{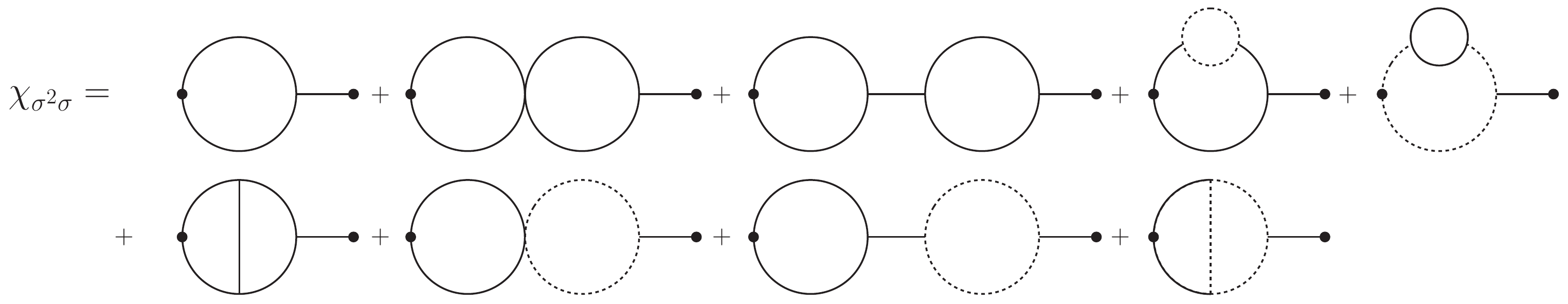}

\includegraphics[scale=0.52]{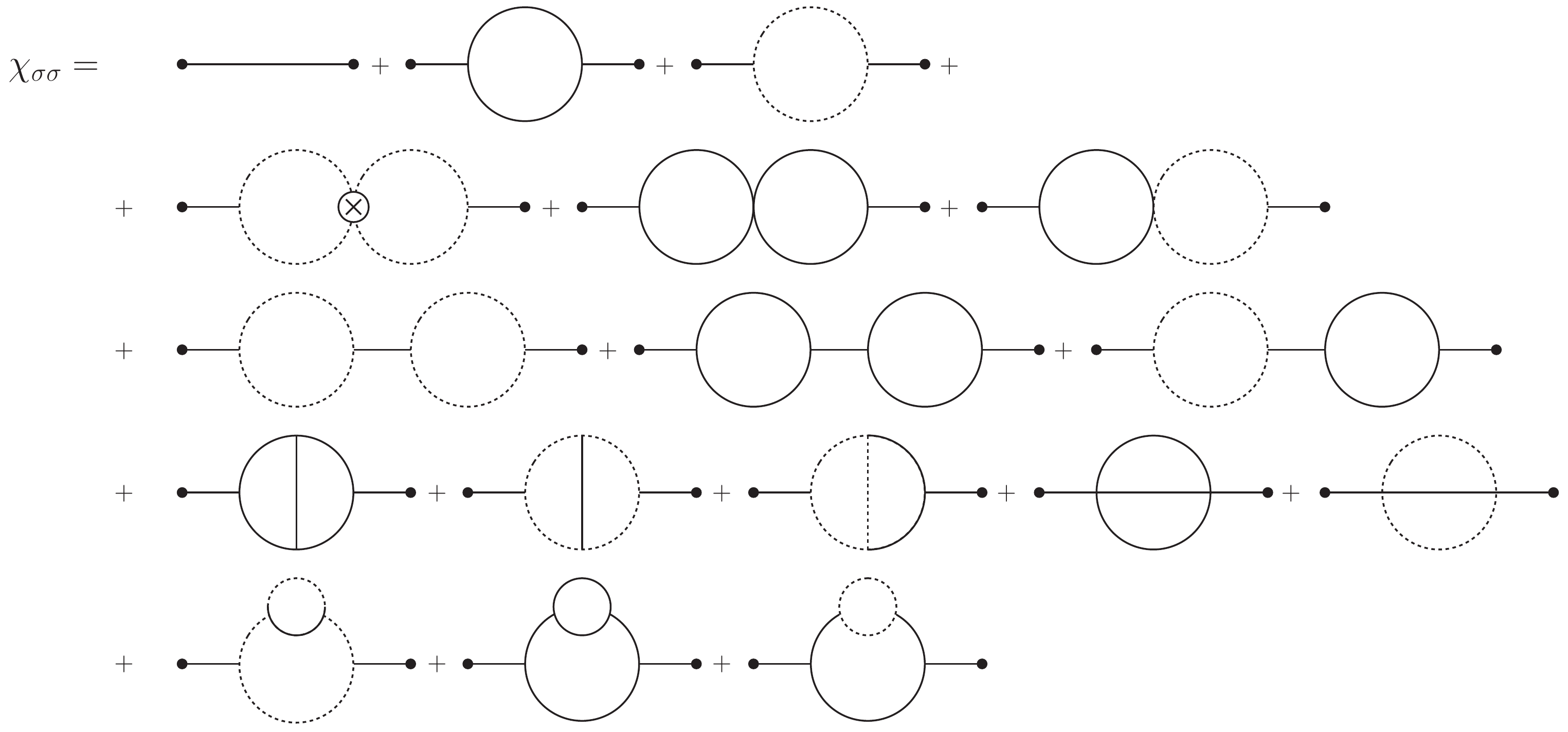}

\protect\caption{\label{fig:Diagramtic-expanssion-of}Diagrammatic expansion in $\varepsilon$,
of the susceptibilities which compose the scalar susceptibility in
the ordered phase. $\chi_{\pi^{2}\pi^{2}},\chi_{\sigma^{2}\sigma^{2}}$
and $\chi_{\pi^{2}\sigma^{2}}$ are calculated to $\mathcal{O}\left(\varepsilon\right)$,
$\chi_{\sigma^{2}\sigma}$ and $\chi_{\pi^{2}\sigma}$ to $\mathcal{O}\left(\varepsilon^{3/2}\right)$
and $\chi_{\sigma\sigma}$ to $\mathcal{O}\left(\varepsilon^{2}\right).$}
\end{figure}

We calculate the scalar susceptibility in the ordered phase to $\mathcal{O}\left(\varepsilon\right)$
by computing the different correlation functions in Eq.\ (\ref{eq:Chi_ro_ro}).
This requires calculation of $\chi_{\pi^{2}\pi^{2}},\chi_{\sigma^{2}\sigma^{2}}$
and $\chi_{\pi^{2}\sigma^{2}}$ to $\mathcal{O}\left(\varepsilon\right)$,
$\chi_{\pi^{2}\sigma}$ and $\chi_{\sigma^{2}\sigma}$ to $\mathcal{O}\left(\varepsilon^{3/2}\right)$,
and $\chi_{\sigma\sigma}$ to $\mathcal{O}\left(\varepsilon^{2}\right)$.
The full diagrammatic expansion of this procedure is presented in
Fig.\ \ref{fig:Diagramtic-expanssion-of}, where our notations for
the Feynman diagrams is shown in Fig.\ \ref{fig:The-standard-notations}.
We find
\begin{eqnarray}
\chi_{\pi^{2}\pi^{2}} & = & \left(N-1\right)\left\{ 2\Pi\left(p,0\right)+U_{c}\left(N-1\right)\left[2E\left(p,0\right)+4L\left(p\right)+\left(\frac{\left(N-1\right)m^{2}}{p^{2}+m^{2}}-N-1\right)\Pi\left(p,0\right)^{2}\right]\right\} ,\label{eq:Chi_p2_p2}
\end{eqnarray}
\begin{eqnarray}
\chi_{\sigma^{2}\sigma^{2}} & = & 2\Pi\left(p,m\right)+2U_{c}\left[-\frac{3}{2}\frac{p^{2}-2m^{2}}{p^{2}+m^{2}}\Pi\left(p,m\right)^{2}+9\left(E\left(p,m\right)+F\left(p,m\right)\right)+\left(N-1\right)F\left(p,0\right)\right],\label{eq:Chi_s2_s2}
\end{eqnarray}
\begin{eqnarray}
\chi_{\pi^{2}\sigma^{2}} & =2U_{c}\left(N-1\right) & \left[-\frac{p^{2}-2m^{2}}{p^{2}+m^{2}}\Pi\left(p,m\right)\Pi\left(p,0\right)+2G\left(p\right)\right],\label{eq:Chi_p2_s2}
\end{eqnarray}
\begin{eqnarray}
4\frac{m}{\sqrt{U_{c}}}\chi_{\sigma^{2}\sigma} & = & -\frac{12m^{2}}{p^{2}+m^{2}}\Pi\left(p,m\right)+U_{c}\frac{4m^{2}}{p^{2}+m^{2}}\left[27\left(E\left(p,m\right)+F\left(p,m\right)\right)-\left(N-1\right)\left(E\left(p,0\right)+3F\left(p,0\right)\right)\right]\label{eq:Chi_s2_s}\\
 &  & +2U_{c}\frac{m^{2}\left(p^{2}-2m^{2}\right)}{\left(p^{2}+m^{2}\right)^{2}}\left[\left(N-1\right)\Pi\left(p,0\right)+9\Pi\left(p,m\right)\right]\Pi\left(p,m\right),\nonumber 
\end{eqnarray}
\begin{eqnarray}
4\frac{m}{\sqrt{U_{c}}}\chi_{\pi^{2}\sigma} & = & 2\left(N-1\right)m^{2}\left[\left(\frac{-2}{p^{2}+m^{2}}+U_{c}\frac{\left(N+1\right)p^{2}+2m^{2}}{\left(p^{2}+m^{2}\right)^{2}}\Pi\left(p,0\right)+3U_{c}\frac{p^{2}-2m^{2}}{\left(p^{2}+m^{2}\right)^{2}}\Pi\left(p,m\right)\right)\Pi\left(p,0\right)\right.\label{eq:Chi_pi_2_s}\\
 &  & \left.-U_{c}\frac{2\left(N-1\right)}{p^{2}+m^{2}}\left(E\left(p,0\right)+2L\left(p\right)+3G\left(p\right)\right)\right],\nonumber 
\end{eqnarray}
\begin{eqnarray}
4\frac{m^{2}}{U_{c}}\chi_{\sigma\sigma} & = & \frac{4}{U_{c}}\frac{m^{2}}{p^{2}+m^{2}}+\frac{m^{4}}{\left(p^{2}+m^{2}\right)^{2}}\left\{ 2\left(N-1\right)\Pi\left(p,0\right)+18\Pi\left(p,m\right)+U_{c}\left[162\left(E\left(p,m\right)+F\left(p,m\right)\right)\right.\right.\label{eq:Chi_s_s}\\
 &  & 3\frac{p^{2}-2m^{2}}{p^{2}+m^{2}}\left(9\Pi\left(p,m\right)^{2}+\left(N-1\right)\Pi\left(p,m\right)\Pi\left(p,0\right)\right)-\frac{\left(N^{2}-1\right)p^{2}+2\left(N-1\right)m^{2}}{\left(p^{2}+m^{2}\right)}\Pi\left(p,0\right)^{2}\nonumber \\
 &  & \left.\left.+6K\left(p,m\right)+2\left(N-1\right)\left(E\left(p,0\right)+6G\left(p\right)+2L\left(p\right)+9F\left(p,0\right)+K\left(p,0\right)\right)\right]\right\} ,\nonumber 
\end{eqnarray}
where
\begin{equation}
\Pi\left(p,0\right)=K_{4-\varepsilon}\left[\frac{1}{2}\left(1+\ln\frac{\Lambda^{2}}{p^{2}}\right)+\frac{1}{8}\varepsilon\left(1+\ln\frac{\Lambda^{2}}{p^{2}}\right)^{2}+\frac{3}{8}\varepsilon\right]+\mathcal{O}\left(\varepsilon^{2}\right),\label{eq:Pi=00005Bp,0=00005D}
\end{equation}
and
\begin{equation}
\begin{array}{ccl}
E\left(p,m_{0}\right) & = & m^{2}\int_{k,q}\frac{1}{q^{2}+m_{0}^{2}}\frac{1}{\left(\mathbf{q}+\mathbf{p}\right)^{2}+m_{0}^{2}}\frac{1}{k^{2}+m_{0}^{2}}\frac{1}{\left(\mathbf{k}+\mathbf{p}\right)^{2}+m_{0}^{2}}\frac{1}{\left(\mathbf{k}+\mathbf{q}\right)^{2}+m^{2}},\\
G\left(p\right) & = & m^{2}\int_{k,q}\frac{1}{q^{2}}\frac{1}{\left(\mathbf{q}+\mathbf{p}\right)^{2}}\frac{1}{\left(\mathbf{k}+\mathbf{p}\right)^{2}+m^{2}}\frac{1}{\left(\mathbf{k-q}\right)^{2}}\frac{1}{k^{2}+m^{2}},\\
L\left(p\right) & = & m^{2}\int_{k,q}\frac{1}{q^{4}}\frac{1}{\left(\mathbf{q}+\mathbf{p}\right)^{2}}\frac{1}{\left(\mathbf{k}+\mathbf{q}\right)^{2}}\frac{1}{k^{2}+m^{2}}\\
F\left(p,m_{0}\right) & = & m^{2}\int_{k}\frac{\Pi\left(k,m_{0}\right)}{\left(k^{2}+m^{2}\right)^{2}}\frac{1}{\left(\mathbf{k}+\mathbf{p}\right)^{2}+m^{2}},\\
K\left(p,m_{0}\right) & = & \frac{1}{m^{2}}\int_{k}\frac{\Pi\left(k,m_{0}\right)}{\left(\mathbf{k}+\mathbf{p}\right)^{2}+m^{2}}.
\end{array}\label{eq:Integrals}
\end{equation}
The integrals $E\left(p,m\right)$ and $G\left(p\right)$ are independent
of the UV cutoff. The remaining terms can be written as 
\begin{equation}
\begin{array}{ccl}
L\left(p\right) & = & K_{4-\varepsilon}^{2}\left[\frac{m^{2}}{4p^{2}}\left(1+\ln\frac{\Lambda^{2}}{m^{2}}\right)+\frac{m^{2}}{4p^{2}}\ln\frac{p^{2}}{p_{0}^{2}}\ln\frac{\Lambda^{2}}{m^{2}}+\Delta L\right],\\
F\left(p,m\right) & = & K_{4-\varepsilon}^{2}\left[\frac{m^{2}}{p^{2}+m^{2}}\frac{\tanh^{-1}x_{m}}{2x_{m}}\left(1+\ln\frac{\Lambda^{2}}{m^{2}}\right)+\Delta F_{1}\right],\\
F\left(p,0\right) & = & K_{4-\varepsilon}^{2}\left[\frac{m^{2}}{p^{2}+m^{2}}\frac{\tanh^{-1}x_{m}}{2x_{m}}\left(1+\ln\frac{\Lambda^{2}}{m^{2}}\right)+\Delta F_{2}\right],\\
K\left(p,m\right) & = & K_{4-\varepsilon}^{2}\left[\frac{2\Lambda^{2}}{m^{2}}+\frac{p^{2}-m^{2}}{2m^{2}}\left(1+\ln\frac{\Lambda^{2}}{m^{2}}\right)^{2}+\frac{m^{2}-2p^{2}}{m^{2}}\left(1+\ln\frac{\Lambda^{2}}{m^{2}}\right)+\Delta K_{1}\right]\\
K\left(p,0\right) & = & K_{4-\varepsilon}^{2}\left[\frac{2\Lambda^{2}}{m^{2}}+\frac{p^{2}-m^{2}}{2m^{2}}\left(1+\ln\frac{\Lambda^{2}}{m^{2}}\right)^{2}+\frac{m^{2}-2p^{2}}{m^{2}}\left(1+\ln\frac{\Lambda^{2}}{m^{2}}\right)+\Delta K_{2}\right]
\end{array}\label{eq:Integrals-1}
\end{equation}
where $p_{0}$ is an IR cutoff on momentum, introduced to regulate
$L(p)$. The terms $\Delta L,\Delta F_{1},\Delta F_{2}$, $\Delta K_{1}$
and $\Delta K_{2}$ are independent of the UV cutoff and are given
by
\[
\Delta L=\frac{m^{2}}{4p^{2}}\text{Li}_{2}\left(-\frac{p^{2}}{m^{2}}\right)-\frac{3m^{2}}{8p^{2}}+\frac{1}{8}\ln\frac{p^{2}}{m^{2}+p^{2}}-\frac{m^{4}}{8p^{4}}\ln\frac{m^{2}}{m^{2}+p^{2}}
\]
\[
\Delta F_{1}=\frac{1}{K_{4-\varepsilon}}\int_{q}\frac{\frac{\sqrt{q^{2}+4m^{2}}}{q}\tanh^{-1}\frac{q}{\sqrt{q^{2}+4m^{2}}}}{\left((\mathbf{p}+\mathbf{q})^{2}+m^{2}\right)\left(q^{2}+m^{2}\right)^{2}},
\]
\[
\Delta F_{2}=\frac{1}{K_{4-\varepsilon}}\int_{q}\frac{\frac{\sqrt{q^{2}+4m^{2}}}{q}\ln\frac{m^{2}}{q^{2}}}{\left((\mathbf{p}+\mathbf{q})^{2}+m^{2}\right)\left(q^{2}+m^{2}\right)^{2}},
\]
\[
\Delta K_{1}=-\frac{2}{K_{4-\varepsilon}}\int_{q}\frac{\left(\frac{\sqrt{q^{2}+4m^{2}}}{q}\tanh^{-1}\frac{q}{\sqrt{4m^{2}+q^{2}}}-\frac{1}{2}\ln\frac{m^{2}}{q^{2}}\right)}{(\mathbf{p}+\mathbf{q})^{2}+m^{2}},
\]
\[
\Delta K_{2}=\frac{m^{2}-p^{2}}{2m^{2}}\left(1-2\text{Li}_{2}\left(\frac{m^{2}}{p^{2}}\right)-\ln^{2}\frac{p^{2}}{m^{2}}+\frac{\pi^{2}}{6}\right)+\frac{2p^{2}}{m^{2}}\ln\frac{2p^{2}}{m^{2}}+2\ln2-2.
\]
Note that the term $L\left(p\right)$ is IR divergent. This can be
avoided by including the next order corrections to the counterterm
of $\pi^{2}$. However, this is not necessary here since we only extract
the universal scaling function to order $\varepsilon^{0}$, and therefore
we are only interested in the UV divergences of the $L\left(p\right)$
term. 

Summation of Eqs.\ (\ref{eq:Chi_p2_p2})-(\ref{eq:Chi_s_s}) yields
a formal expression for the scalar susceptibility to $\mathcal{O}\left(\varepsilon\right)$,

\begin{eqnarray}
\chi_{s}\left(p\right) & = & \frac{4}{U_{c}}\frac{m^{2}}{p^{2}+m^{2}}+2\frac{\left(p^{2}-2m^{2}\right)^{2}}{\left(p^{2}+m^{2}\right)^{2}}\left\{ \Pi\left(p,m\right)-\frac{3}{2}U_{c}\frac{p^{2}-2m^{2}}{p^{2}+m^{2}}\Pi\left(p,m\right)^{2}-U_{c}\frac{\left(N-1\right))p^{2}}{p^{2}+m^{2}}\Pi\left(p,m\right)\Pi\left(p,0\right)\right.\label{eq:SG Skeleton}\\
 &  & \left.+U_{c}\left(9\left(E\left(p,m\right)+F\left(p,m\right)\right)+\left(N-1\right)F\left(p,0\right)\right)\right\} +U_{c}\frac{2m^{6}}{\left(p^{2}+m^{2}\right)^{3}}\left(3K\left(p,m\right)+\left(N-1\right)K\left(p,0\right)\right)\nonumber \\
 &  & +\frac{2\left(N-1\right)p^{4}}{\left(p^{2}+m^{2}\right)^{2}}\left\{ \Pi\left(p,0\right)-U_{c}\left[\frac{2m^{2}+\left(N+1\right)p^{2}}{2\left(p^{2}+m^{2}\right)}\Pi\left(p,0\right)^{2}+2L\left(p\right)+E\left(p,0\right)+2\left(1-2\frac{m^{2}}{p^{2}}\right)G\left(p\right)\right]\right\} .\nonumber 
\end{eqnarray}
We can use this expression to obtain the logarithmic UV divergences
at $\mathcal{O}(\varepsilon)$. From this, using Eq.\ (\ref{eq:dis_expansion}),
we extract the universal scaling function. Indeed, we find that the
result of this analysis yields Eq.\ (\ref{eq:eq:phi tilde ordered}),
and that the regular part $\chi_{reg}$ matches that obtained in the
disordered phase, Eq.\ (\ref{eq:chireg}).

\section{Polarization bubble at finite temperatures\label{appendix:COMPUTATION-OF-}}

We compute $\pi_{0}^{''}$ and $\pi_{1}^{''}$, the imaginary parts
of $\pi_{0}$ and $\pi_{1}$, see Eq.\ (\ref{eq:S=00005Bp=00005D High_T}).
These terms are used in Sec.\ \ref{sec:Expansion-in-Finite} in order
to obtain the universal scaling function in the quantum critical regime.

The sum in Eq.\ (\ref{eq:Pi General High_T}) can be performed by
using the identity\cite{podolsky2011visibility}
\begin{eqnarray}
T\sum_{m}\frac{1}{\omega_{m}^{2}+a^{2}}\frac{1}{\left(\omega_{m}+\omega_{n}\right)^{2}+b^{2}} & = & \frac{n\left(-a\right)}{2a\left(\left(i\omega_{n}-a\right)^{2}-b^{2}\right)}-\frac{n\left(a\right)}{2a\left(\left(i\omega_{n}+a\right)^{2}-b^{2}\right)}+\label{eq:Sum of highT}\\
 &  & \frac{n\left(-b\right)}{2a\left(\left(i\omega_{n}+b\right)^{2}-a^{2}\right)}-\frac{n\left(b\right)}{2a\left(\left(i\omega_{n}-b\right)^{2}-a^{2}\right)}\nonumber 
\end{eqnarray}
where $n\left(\nu\right)$ is the Bose-Einstein occupation function,
\[
n\left(\nu\right)=\frac{1}{\exp\left(\nu/T\right)-1}.
\]
We insert Eq.\ (\ref{eq:Sum of highT}) into Eq.\ (\ref{eq:Pi General High_T})
to obtain
\begin{equation}
\Pi_{T}^{0}\left(\omega_{n}\right)=K_{4-\varepsilon}\int_{q}\frac{1}{\sqrt{q^{2}+m_{T}^{2}}}\frac{\coth\frac{\sqrt{q^{2}+m_{T}^{2}}}{2T}}{q^{2}+m_{T}^{2}+\frac{\omega_{n}^{2}}{4}}q^{2-\varepsilon}dq\label{eq:Pi high T explicit}
\end{equation}
We are only interested in the finite part of Eq.\ (\ref{eq:Pi high T explicit})
to $\mathcal{O}\left(\varepsilon^{0}\right)$. We will therefore subtract
the divergent part by writing
\begin{equation}
\pi_{0}\left(\omega_{n}\right)=\int_{0}^{\infty}\frac{1}{\sqrt{q^{2}+m_{T}^{2}}}\left[\frac{\coth\frac{\sqrt{q^{2}+m_{T}^{2}}}{2T}}{q^{2}+\frac{\omega_{n}^{2}}{4}+m_{T}^{2}}-\frac{1}{q^{2}+m_{T}^{2}}\right]q^{2}dq.\label{eq:pi0}
\end{equation}
We have normalized Eq.\ (\ref{eq:pi0}) by $K_{4-\varepsilon}$ in
order for $\pi_{0}\left(\omega_{n}\right)$ to be consistent with
Eq.\ (\ref{eq:P=00005Bp,T=00005D, geneal form}). 

We now replace $\omega_{n}$ with real frequencies by performing a
Wick rotation, $\omega_{n}\rightarrow-i\omega+0^{+}$,

\begin{equation}
\pi_{0}\left(\omega\right)=\int_{0}^{\infty}\frac{1}{\sqrt{q^{2}+m_{T}^{2}}}\left[\frac{\coth\frac{\sqrt{q^{2}+m_{T}^{2}}}{2T}}{q^{2}-\frac{\omega^{2}}{4}-i0^{+}+m_{T}^{2}}{\rm sign}\left(\omega\right)-\frac{1}{q^{2}+m_{T}^{2}}\right]q^{2}dq.\label{eq:Pi0_Integral}
\end{equation}
The imaginary part of Eq.\ (\ref{eq:Pi0_Integral}) can be computed
by using the identity,
\begin{equation}
\frac{1}{x+i0^{+}}=\mathcal{P}\left(\frac{1}{x}\right)-i\pi\delta\left(x\right)\label{eq:Identity Principle}
\end{equation}
where $\mathcal{P}$ denotes the principal value. We insert Eq.\ (\ref{eq:Identity Principle})
into Eq.\ (\ref{eq:Pi0_Integral}) to find
\begin{equation}
\pi_{0}^{''}\left(\omega\right)=\pi\int_{0}^{\infty}\frac{1}{\sqrt{q^{2}+m_{T}^{2}}}\coth\frac{\sqrt{q^{2}+m_{T}^{2}}}{2T}{\rm sign}\left(\omega\right)\delta\left(q^{2}-\frac{\omega^{2}}{4}+m_{T}^{2}\right)q^{2}dq.\label{eq:P0=00005Bp,t=00005D_Final}
\end{equation}
By performing the integral in Eq.\ (\ref{eq:P0=00005Bp,t=00005D_Final}),
we obtain

\begin{equation}
\pi_{0}^{''}\left(\omega\right)=\frac{\pi}{2\left|\omega\right|}\sqrt{\omega^{2}-4m_{T}^{2}}\Theta\left(\left|\omega\right|-2m_{T}\right)\coth\frac{\omega}{4T}.\label{eq:p0}
\end{equation}
We can obtain, in a similar manner, 
\[
\pi_{1}^{''}=-\frac{1}{2}\pi_{0}^{''}\left(\omega\right)\ln\left|\frac{\omega^{2}}{4}-m_{T}^{2}\right|.
\]

\twocolumngrid

\bibliographystyle{apsrev}
\bibliography{Higgs4-e}

\end{document}